\documentclass[12pt,preprint]{aastex}

\newcommand{\msun}{${\rm M}_{\odot}$}
\newcommand{\kms}{${\rm km\,s}^{-1}$}
\newcommand{\sfr}{${\rm M}_{\odot}\,{\rm yr}^{-1}$}

\newcommand{\wb}{$B_{435}$}
\newcommand{\wv}{$V_{606}$}
\newcommand{\wi}{$i_{775}$}
\newcommand{\wz}{$z_{850}$}
\newcommand{\Lya}{Ly${\alpha}$}
\newcommand{\hbl}{${\rm km\,s}^{-1}\,{\rm Mpc}^{-1}$}

\shorttitle{Cold gas accretion onto $z\approx 1.6$ star--forming galaxies}
\shortauthors{Giavalisco et~al.}

\begin{document}


\title{Discovery Of Cold, Pristine Gas Possibly Accreting Onto An Overdensity
  Of Star--Forming Galaxies At Redshift $z\sim 1.6$\altaffilmark{1}.}


\author{Mauro    Giavalisco\altaffilmark{2}, 
        Eros     Vanzella\altaffilmark{3},
        Sara     Salimbeni\altaffilmark{2}, 
        Todd M.  Tripp\altaffilmark{2},
        Mark     Dickinson\altaffilmark{4}, 
        Paolo    Cassata\altaffilmark{2}, 
        Alvio    Renzini\altaffilmark{5}, 
        Yicheng  Guo\altaffilmark{2}, 
        Henry C. Ferguson\altaffilmark{6}, 
        Mario    Nonino\altaffilmark{3},
        Andrea   Cimatti\altaffilmark{7}, 
        Jaron    Kurk\altaffilmark{8},
        Marco    Mignoli\altaffilmark{9} and
        Yuping   Tang\altaffilmark{2} }
\affil{$^{2}$Astronomy Department, University of Massachusetts, Amherst, MA
  01003, U.S.A.}  
\affil{$^{3}$INAF--Osservatorio Astronomico di Trieste, I-40131, Italy}
\affil{$^{4}$National Optical Astronomy Observatories, Tucson, AZ 85719, U.S.A.}
\affil{$^{5}$INAF--Osservatorio Astronomico di Padova, I-35122, Italy}
\affil{$^{6}$Space Telescope Science Institute, Baltimore, MD 21218, U.S.A.}
\affil{$^{7}$Dipartimento di Astronomia, Universit\'a degli Studi di Bologna,
  I-40127 Bologna, Italy}  
\affil{$^{8}$Max Planck Institut f\"ur Extraterrestrische Physik, D-85748
  Garching, Germany} 
\affil{$^{9}$INAF-Osservatorio Astronomico di Bologna, I-40127, Italy}
\email{mauro@astro.umass.edu}

\altaffiltext{1}{Based on observations obtained with European Southern
  Observatories Very Large Telescope, Chile, and with the NASA/ESA {\it Hubble
    Space Telescope}, obtained at the Space Telescope Science Institute, which
  is operated by the Association of Universities for Research in Astronomy,
  Inc. (AURA) under NASA contract NAS 5-26555.}

\begin{abstract}
We report the discovery of large amounts of cold ($T\sim~10^4$ K), chemically
young gas in an overdensity of galaxies at redshift $z\approx 1.6$ located in
the Great Observatories Origins Deep Survey southern field (GOODS--S). 
The gas is identified thanks to the ultra--strong \ion{Mg}{2}~$\lambda2800$ 
absorption features it imprints in the rest--frame UV spectra of galaxies 
in the background of the overdensity. 
There is no evidence that the optically--thick gas is part of any massive
galaxy (i.e. $M_{star}>4\times 10^9$ \msun), but rather is associated with the
overdensity; less massive and fainter galaxies ($25.5<z<27.5$ mag) have too
large an impact parameter to be causing ultra--strong absorption systems,
based on our knowledge of such systems. The lack of corresponding
\ion{Fe}{2}\ absorption features, not detected even in co--added spectra,
suggests that the gas is chemically more pristine than the ISM and outflows of
star--forming galaxies at similar redshift, including the galaxies of the
overdensity itself,  
and comparable to the most metal--poor stars in the Milky Way halo. A
crude estimate of the projected covering factor of the high--column density
gas ($N_H\gtrsim10^{20}$ cm$^{-2}$) based on the observed fraction of galaxies
with ultra--strong absorbers is $C_F\approx 0.04$.
A broad, continuum absorption profile extending to the red of the interstellar
\ion{Mg}{2}\ absorption line by $\lesssim 2000$ \kms\ is possibly detected 
in two independent co--added spectra of galaxies belonging to the overdensity, 
consistent with a large--scale infall motion of the gas onto the overdensity
and its galaxies. Overall, these findings provides the first tentative evidence
of accretion of cold, chemically young gas onto galaxies at high redshift,
possibly feeding their star formation activity. We suggest that the fact that
the galaxies are members of a large structure, as opposed to field galaxies,
might play a significant role in our ability to detect the accreting gas.
\end{abstract}




\section{Introduction}
The history of gas accretion in galaxies is a major outstanding question that
is essentially unconstrained by the empirical investigation. Do galaxies
evolve by transforming into stars the mass of gas present in their hosting
halo at the time of its formation? Or do they continue accreting substantial
amounts of gas from the environment to feed star formation?

Theoretically, an interesting recent development is the realization that most
of the baryonic mass in galaxies of all masses is supposed to accrete as
relatively cold gas, namely gas that was never shock--heated to the halo
virial temperature (e.g. Birnboim \& Dekel 2003; Keres et~al. 2005; Dekel \&
Birnboim 2006; Keres et~al. 2009; Dekel et el. 2009a and references
therein). Current hydro--dynamical simulations show that halos with mass below
the threshold value $M_{thr}\sim~2-3\times 10^{11}$ \msun\ acquire their gas
primarily by cold accretion, with hot accretion remaining relatively
unimportant in providing gas for star formation. More massive halos do acquire
an atmosphere of hot coronal gas in the simulations which subsequently cools
and contributes to star formation, but cold accretion remains nonetheless the
dominant mode of gas acquisition, even in this case, at high redshift
(e.g. $z\gtrsim 2$), while hot accretion starts to compete only at $z<1$.

Observationally, there has been scant evidence for galactic-scale accretion so
far, cold or otherwise, with powerful and ubiquitous outflows powered by star
formation being the only bulk--motions of gas in the proximity of galaxies
that have been unambiguously detected at low and high redshift (e.g. Steidel
et~al. 1996; Franx 1997; Pettini et~al. 2001; Shapley et~al. 2003; Tremonti
et~al. 2007; Weiner et~al. 2009; Nestor et~al. 2010). Rauch et~al. (2008)
reported detection of Lyman--$\alpha$ emission around galaxies at $z\sim 3$
that they interpret as possible evidence of accreting \ion{H}{1}\ gas onto
these systems. Steidel et~al. (2010) find evidence of possible gas inflows
around relatively massive galaxies at $z\sim~2$ but conclude that this is not
necessarily evidence of cold accretion and other explanations are possible,
and, in fact, more likely. Indirect evidence of accretion of chemically
pristine gas is reported by Cresci et al. (2010), who interpret the inverse
gradient of the metallicity of the nebular gas in star--forming galaxies at
$z\sim 3$ as evidence that the metallicity of the central regions is being
diluted by accreting cold flows.

It is not straightforward, however, to interpret the absence of evidence of
accretion as evidence of the absence of accretion. On the one hand, we do not
have much insight on the relative covering factor of high--column density
inflowing gas, i.e.  observable in relatively low S/N absorption systems,
versus the outflowing one. For example, the simulations predict the geometry
of the cold gas to be largely filamentary over scales ranging from $10^2$ kpc
to $\sim$Mpc (Dekel et~al. 2009a), in particular before they enter the
``sphere of influence'' of the accreting galaxy (the circum--galactic medium,
or CGM, in recent parlance). At closer galactocentric distance, which is where
current observations are probing the ISM, the smooth accreting flow is
expected to break into a lumpier distribution of cold gas, especially around
more massive halos (Keres \& Hernquist 2009; Dekel et al 2009b), that in turn
feeds clumpy, unstable disks. If this is a fair description of reality, then
the total cross section of high--column density accreting gas, especially in
the proximity of the star--forming regions, should be relatively small, as
suggested in recent works (Fumagalli et~al. 2011; Ceverino et~al. 2010; Kimm
et~al. 2010; Dekel, private communication), which calculate that the covering
factor of gas with $N_H<10^{20}$ cm$^{-2}$ around $M\sim 10^{12}$ \msun\ halos
is $C_F\lesssim 3$\%, and that of gas with $10^{20}<N_H<10^{21}$ cm$^{-2}$ is
$C_F\lesssim 1$\%.

On the other hand, there is the distinct possibility that we do not fully
understand the inherent bias and limitations of the current observations of
the circum--galactic medium of distant galaxies, as far as the detection of
gas accretion is concerned. For example, the observations by Steidel et
al. (2010) probe the gas along sight lines toward, and in some cases in the
proximity ($\lesssim 150$ kpc) of, regions of powerful star formation, with
the UV--bright regions of the galaxies providing the back--illumination to
study the gas in absorption. The galaxies are selected to be star--forming
ones with moderate or no dust obscuration. These selection criteria might
preferentially yield samples of galaxies oriented along the {\it least
  obscured line sight} from the observer toward their center, thus biasing
their studies to finding strong signature of outflows with high covering
factor.

In general, outflows from star--forming galaxies cover large solid angles
(Tremonti et~al. 2007; Martin 2006), and it is possible that their
interactions with the inflowing gas, as well as the sensitivity of the
observations, conspire against the detectability of inflows, because the
strength of the interstellar absorption is generally very strong and broad in
velocity space, and can completely dilute and hide the signal of the inflow
(Kimm et~al. 2010). Unfortunately, the simulations do not yet provide firm
predictions on the observability (with current technology) of the inflowing of
gas and how this interacts with the outflowing one, especially in the volumes
at small impact parameters from the galaxies' centers that have been targeted
by the observations so far.

Until we gain firmer theoretical guidance on the effective cross section of
the accreting gas and its observability in the proximity of galaxies, an
important first step would be to constrain the abundance of gas at large
galactocentric distance that can plausibly feed accretion. Indeed, staying
well clear of the regions directly affected by the outflows seems a promising
strategy, since the gas must reach the accreting galaxies from large
distances, and in this way one would eliminate the inherent complications of
having to disentangle and interpret the kinematics of gas that participates in
inflows as well as outflows. The challenge of such a strategy is to overcome
the small cross section of high--column density gas, if this is to be detected
by means of absorption spectroscopy. This might appear to be a daunting task,
especially if the theory is right that, at large separations from galaxies,
the accreting flows exhibit filamentary geometry. Cosmological simulations,
however, suggest, at least qualitatively, that the projected cross section of
gas accreting onto large cosmic structures is proportionally larger than in
the case of isolated galaxies. The simulations also show that the covering
fraction of the gas increases with the halo mass. Thus, targeting the
relatively common ``overdensities'' of galaxies, nearly ubiquitous in any
redshift survey as ``spikes'' in the galaxies' redshift distribution, might
currently be our best hope to gain some insight into the issue of ``cold
accretion'' since they might provide sufficient cross section of high--column
density gas. The most massive among these overdensities typically extend over
wide regions of space, with linear size of a few tens Mpc, which means that a
relatively large number of bright background galaxies are generally available
to back--illuminate the gas and make it visible by means of spectroscopy of
the intervening absorption systems.

In this paper we report the discovery of large amounts of cold gas,
i.e. $T\sim~ 10^4$ K, not directly associated with any galaxies, in a large
overdensity of galaxies at $z\approx 1.6$ located in the GOODS--South field
(see Kurk et~al. 2009). The gas is detected thanks to the
\ion{Mg}{2}~$\lambda\lambda2797,2803$ absorption features it imprints in the
individual spectra of background Lyman--break galaxies at $z\sim~3$, as well
in the stacked spectra of other background galaxies. It appears to be
chemically more pristine than that observed in the outflows of star--forming
galaxies at the same redshift, including those within the same overdensity. We
discuss the observed and inferred properties of the gas and the tentative
evidence that it accretes onto at least some of the overdensity galaxies.
Throughout this paper, when needed, we use a CDM world model with Hubble
constant $H_0=70$ \hbl, $\Omega_m=0.3$ and $\Omega_{\Lambda}=0.7$. Magnitudes
are in the AB scale of Oke \& Gunn (1977).

\section{Observations and Data Set}

The data presented in this paper include spectra of star--forming galaxies in
the GOODS--S field, selected from the GOODS {\it HST}/ACS imaging survey
(Giavalisco et~al. 2004a) using a number of criteria based on rest--frame UV
colors aimed at selecting star--forming galaxies at $z>1.5$ (see Giavalisco
et~al. 2004b; Vanzella et~al. 2005, 2006, 2008). The ACS mosaic in the
GOODS--S field covers an area of roughly $10'\times 16'$, or about $5\times 8$
Mpc (proper) at $z=1.6$, and has been the target of a number of programs of
follow--up spectroscopic observations of {\it HST}--selected galaxies using a
number of telescope and spectrometer combinations, including VLT with VIMOS
and FORS2 (Vanzella et~al. 2005, 2006, 2008; Popesso et~al. 2009; Balestra
et~al. 2010) and Keck with DEIMOS and LRIS (Stern et~al., in preparation ); in
this work we focus on the spectra obtained with FORS2, which we will refer to
as the ``GOODS survey''. These include three Lyman--break galaxies (LBGs) at
$z\gtrsim 3$ where we have discovered absorption features, the subject of our
discussion here, that we identify as due to \ion{Mg}{2}\ at $z\approx
1.6$. These spectra were obtained as part of our program of spectroscopic
identifications of Lyman--break galaxy candidates selected from the ACS
\wb\wv\wi\wz\ images (Giavalisco et~al. 2004b; Vanzella et~al. 2009). 
Acquisition, reduction and analysis of the spectra have already been presented
in the papers above, and we will not discuss them further here. 

Other spectroscopic data include the spectra of star--forming galaxies at
$1.5<z<2.5$ obtained during the ``Galaxy Mass Assembly ultra--deep
Spectroscopic Survey'' (GMASS) survey (Cimatti et~al. 2008; Kurk et~al. 2009),
which have been selected from the GOODS Spitzer/IRAC 4.5 \micron\ images to
satisfy the flux limit $m_{4.5\mu}\le 23$, essentially a mass--limit selection
criterion approximately corresponding to $m_{star}>2\times 10^{9}$ \msun, with
a small dependence on the spectral type. The redshift of these galaxies
extends up to $z<5.5$ with $\approx 90$\% of them having $z<2.5$, and include
the redshifts of galaxies that belong to a large overdensity at $z\approx
1.61$, which are distributed over a substantial portion of the ACS mosaic, as
we will discuss later (Vanzella et~al. 2008; Kurk et~al. 2009). In the
following, we will refer to these redshift as ``the GMASS survey''. Including
GMASS and GOODS, there are 97 spectroscopic redshifts and 194 photometric
redshifts (Dahlen et~al. 2010) in the range $1.56\le z\le 1.64$, which roughly
defines the overdensity, satisfying the above mid--IR flux limit criterion,
crudely $\approx 50$\% spectroscopic completeness in that redshift interval.

\begin{figure}
\epsscale{0.85}
\plotone{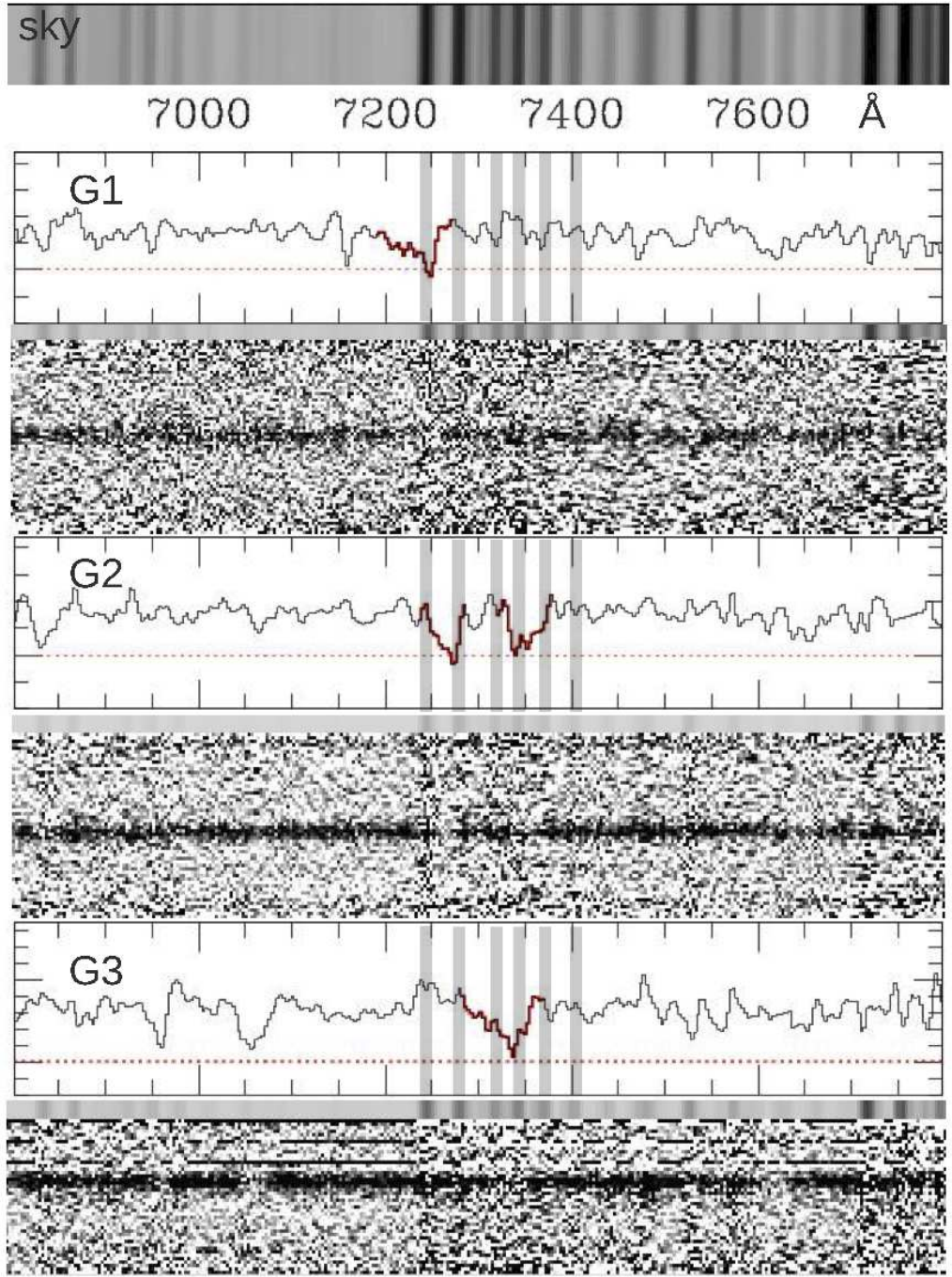}
\caption{The observed spectra of the three Lyman--break at $z>3$ with
  intervening absorption system at $z\approx 1.6$, which we identify as due to
  \ion{Mg}{2}\ (highlighted). The 2--D dispersed images have been
  sky--subtracted and wavelength--calibrated (in the observed frame). A
  boxcar--smoothed, optimally--extracted spectrum is also shown on top of the
  2--D spectrum together with the adjacent spectrum of the night sky included
  in the same slitlet. A spectrum of the night sky is shown at the top of the
  figure, and major night sky emission line are marked as shaded vertical
  bands.
\label{fig:lbgabs_2D}}
\end{figure}

\begin{figure}
\epsscale{1.0}
\plotone{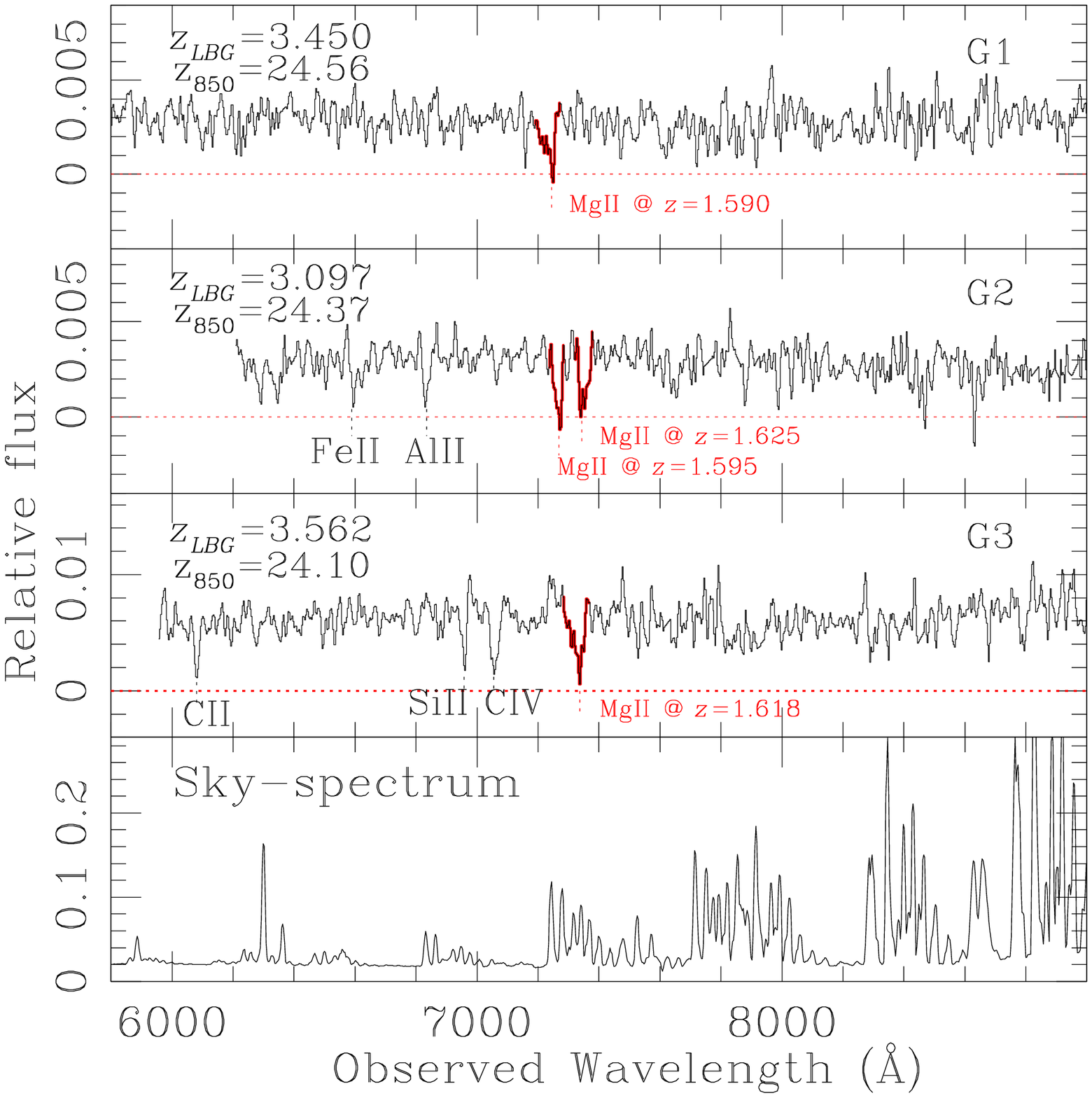}
\caption{The optimally--extracted spectra of three Lyman--break at $z>3$ with
  intervening absorption systems at $z\approx 1.6$ that we identify as due to
  \ion{Mg}{2}. The \wz\ magnitude and spectroscopic redshift of the
  LBG are shown, together with the redshift of the absorber, derived from the
  central wavelength of the line and average wavelength of the
  \ion{Mg}{2}\ doublet, $\lambda=2800$ \AA. Also shown are some spectroscopic
  feature at the redshift of the Lyman--break galaxies. Galaxy G1 is an
  example of LBG with very weak interstellar absorption lines but strong
  Lyman-$\alpha$ emission line (not shown); G2 and G3 are cases where the
  absorption lines are stronger, but the Lyman-$\alpha$ weaker or absent (see
  text and Vanzella et~al. 2009). The night sky spectrum at the bottom is from
  the same mask used to observe the galaxies.
\label{fig:lbgabs_1D}}
\end{figure}

Figure \ref{fig:lbgabs_2D} shows the observed 2--D GOODS spectra of three
Lyman--break galaxies, out of a total of 21 with $z_{850}<24.6$ that we have
analyzed, in which we have detected the presence of absorption features, all
of the them around $\lambda\sim 7300$ \AA, that do not match any known strong
feature at their redshift. These Lyman--break galaxies have originally been
selected as ``B--band'' dropouts from the GOODS {\it HST}/ACS
\wb\wv\wi\wz\ photometry (Giavalisco et~al. 2004b) and marked for follow--up
spectroscopy as part of our investigation of galaxies at very high redshift
(see Vanzella et~al. 2008, 2009 for a discussion of the data reduction and
analysis). Figure \ref{fig:lbgabs_1D} shows the optimally--extracted spectra
of the three LBG with the intervening absorption features marked in red and
labeled with the redshift of the trough, if identified as \ion{Mg}{2}. The
\wz\ magnitude and redshift of the LBG are also shown, as well as some of
their spectroscopic feature.

The dispersion of these spectra is $R\approx 660$, corresponding to a
resolution of $\approx 11$ \AA, or about 454 \kms\ in the observed frame, at
the wavelength of the features. The pixel scale along the dispersion direction
is $\Delta\lambda=3.2$ \AA.

While all four absorption features have wavelength within the range of the
band of weak night sky lines located at $7200\lesssim\lambda\lesssim7400$ \AA,
there is no evidence that the features themselves might be due to noisy sky
removal. Both the 2--D and extracted spectra show that no similar features are
observed in spectral regions affected by much stronger night sky lines, where
the continuum of the three LBG is observed without spurious ``absorption
lines.'' In general, the four LBG absorption features have FWHM consistent
with kinematic broadening being comparable to the spectral resolution, except
the absorption feature in the spectrum of G3 (see below), which is
considerably broader.

Galaxy J033220.85-275038.9 (G1 from now on) has apparent magnitude
$z_{850}=24.56$ mag and redshift $z=3.450$, measured from a spectrum with
quality flag $QF=B$ in the quality classification system by \citet{van09}. The
quality flag values A, B and C, correspond to redshift identifications that
are ``unambiguous'', ``likely'' e.g. when based on one line only or on a
spectral break, and ``uncertain'', respectively. As discussed in \citet{van09}
and \cite{van08}, we assigned QF=B\ to G1 because only one strong emission
line, interpreted as Lyman--$\alpha$, is observed in its spectrum; the
asymmetric profile of the line, the dimming of the continuum in correspondence
of the Lyman-$\alpha$ forest and the presence of a strong spectral break in an
otherwise blue continuum, as shown by its ACS broad--band colors, make the
redshift identification of this galaxy secure. An absorption line is detected
with central wavelength located at $\lambda=7246.5\pm 0.5$ \AA, observed
equivalent width $W_o=20\pm 4$ \AA\ and ${\rm FWHM}=12\pm 2$ \AA\ (from a fit
to a single gaussian line profile). Although the spectrum has relatively low
S/N, the 2--D spectrum shows that the absorption feature is broader than the
night sky line adjacent to it. The extracted spectrum suggests that the line
possibly has an asymmetric profile, with the blue wing broader than the red
one.

Galaxy J033226.18-275211.3 (or G2) has magnitude $z_{850}=24.37$ mag, and in
\citet{van09}, we reported its redshift measure as having $QF=C$ and listed
the tentative value $z=3.097$. Since this source shows two absorption systems
at around $\lambda\approx 7280$, we have re-analyzed both our FORS2 spectrum
and the VIMOS spectrum of this galaxy acquired by Popesso et~al. (2009) in an
attempt to make a better measure of its redshift. Using the spectrum of cB58,
a lensed LBG whose apparent magnitude is unusually bright (Pettini
et~al. 2003), as a template, a cross--correlation analysis yields $z_F=3.0950$
from the FORS2 spectrum and $z_V=3.0954$ from the VIMOS one. In the former
case, the redshift is mostly determined by \ion{C}{4}, \ion{Fe}{2}\ and
\ion{Al}{2}\ absorption lines (these two visible in the portion of the
spectrum plotted in Figure \ref{fig:lbgabs_1D}), while in the latter by \Lya,
\ion{C}{2}, and \ion{C}{4}\ features. Two absorption lines inconsistent with
known strong features at $z\sim~3$ are observed with central wavelength at
$\lambda_c=7266.2$ and $7352.1$ \AA, the latter possibly showing evidence of a
barely resolved doublet. The 2--D spectrum shows that in both cases the
absorption feature is broader than adjacent night sky lines. The observed
equivalent width of the features is $W_o=23\pm 5$ and $W_o=14\pm 5$ \AA,
respectively, while the width from best fits to single-gaussian line profiles
are ${\rm FWHM}=21\pm 2$ \AA\ and $24\pm 2$ \AA, respectively. The velocity
difference between the two absorption lines is $\approx 3500$ \kms, too large
to believe that one intervening galactic system (a galaxy and its circum
galactic medium) might be responsible for producing both of them.

Galaxy J033226.76-275225.9 (G3), with $z_{850}=24.10$ mag and redshift
$z=3.562$, has quality flag QF=A. Some of the low--ionization metal absorption
lines used to determine its redshift are visible in the figure. An absorption
line is observed with central wavelength located at $\lambda=7331.6$ \AA\ and
with equivalent width and line width $W_o=26\pm 7$ and ${\rm FWHM}=45\pm 10$
\AA, respectively (again, the width comes from a best fit to a single gaussian
line profile). As in the case of G1, it is possible that the profile of this
line is asymmetric, with the bluer part broader than the red one. Finally,
also in this case the 2--D spectrum does not show any evidence of an anomalous
subtraction of the sky emission, and the feature appears to be significantly
broader than the night sky emission lines close to it.

The three Lyman--break galaxies are fairly typical in their variety of
spectral properties. Galaxy G1 has strong Lyman--$\alpha$ in emission and weak
interstellar absorption lines, while G2 and G3 have stronger interstellar
absorption features and weak or absent Lyman--$\alpha$, as very frequently
observed in what are termed ``emitters'' and ``absorbers'' LBGs (see
\citet{s96}; \citet{sha03}; \citet{van09} and discussion therein). On the
plane of the sky the three LBGs are roughly aligned along a straight line,
oriented approximately at ${\rm PA}=-45$\degr; the angular separations between
the centroids of the \wz--band images of G1 and G2, G1 and G3, and G2 and G3
are 116.3, 132.6 and 16.5 arcsec, respectively (see Figure
\ref{fig:overdens_dahlen} and \ref{fig:overdens_guo}). At redshift $z=1.61$
the corresponding projected distances between the line--of--sight (LOS) to G1
and G2, G1 and G3 and G2 and G3 are $d_{12}=986$, $d_{13}=1,124$ and
$d_{23}=140$ kpc (physical), respectively. In velocity space, the absorption
feature of G1 and the bluer of the two features of G2 are separated by $\Delta
V=-580$ \kms, while the redder one and the feature of G3 have $\Delta V=800$
\kms. The separation between the absorption lines of G1 and G3 is $-3520$
\kms, the same as that the velocity difference between the two features
observed in the spectrum of G2.

\begin{figure}
\epsscale{0.9}
\plotone{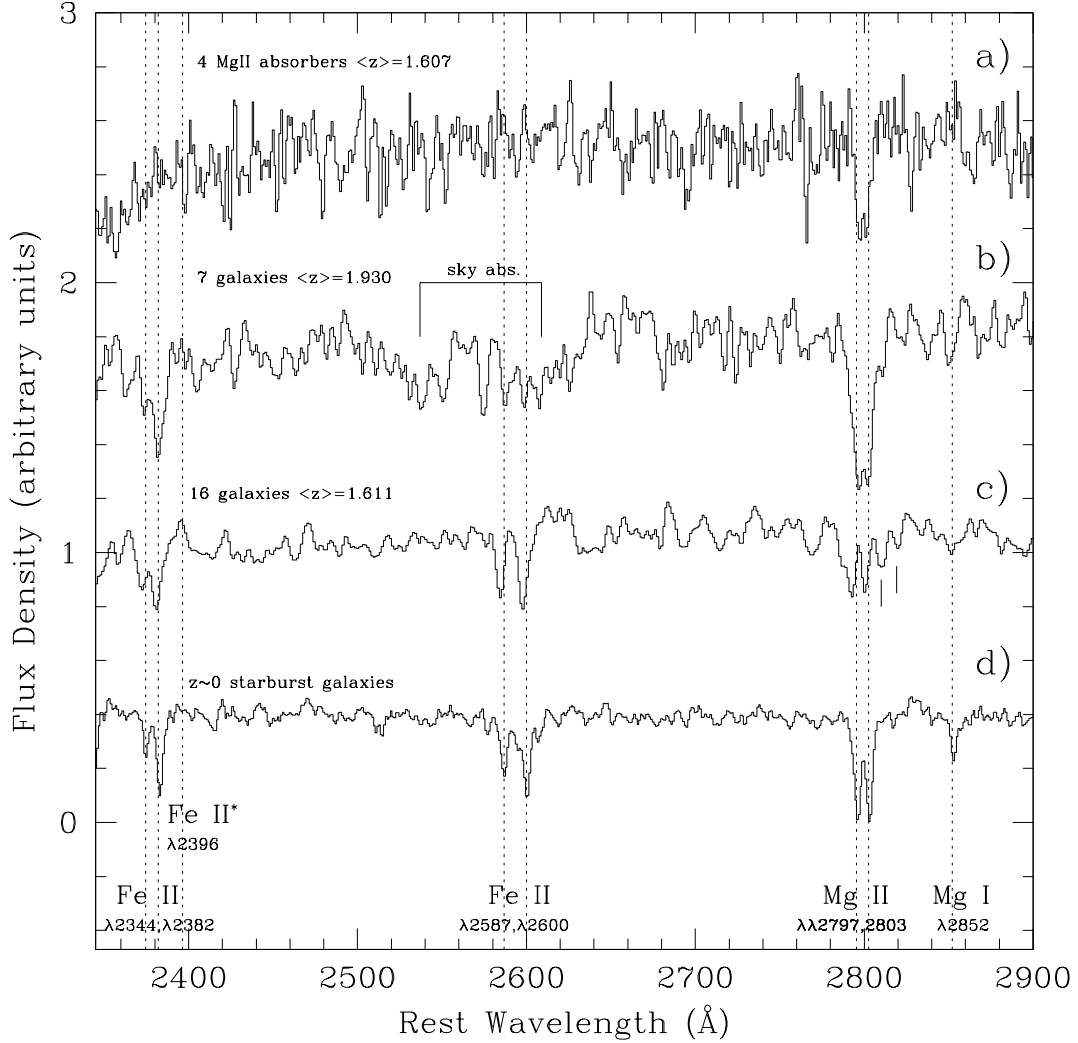}
\caption{Co--added spectra from the GOODS survey. From top to bottom: a) The
  stack of the four \ion{Mg}{2}\ absorption lines shown in Figure
  \ref{fig:lbgabs_1D}, after shifting the central wavelength of each line to
  the common rest--frame wavelength $\lambda=2800$ \AA. A partially resolved
  doublet is clearly observed, with the centers of the blue and red component
  located at $\lambda=2796.7$ \AA\ and $\lambda=2801.5$ \AA, respectively, and
  total rest equivalent width $W_r=6.3\pm 1.5$ \AA. No other absorption
  feature is observed in the stacked spectrum down to a 1--$\sigma$ equivalent
  width limit of $1.5$ \AA; b) a stack of seven spectra of star--forming
  galaxies at $z\sim 1.9$, shown here for comparison because it has only
  marginally better S/N than that of the four LBGs. Absorption lines of
  \ion{Mg}{2}\ and \ion{Fe}{2}\ are observed, although the features around
  $\lambda\approx2600$ \AA\ are severely affected by telluric absorption. The
  profile of the \ion{Mg}{2}\ absorption line is very similar to that of the
  four LBG absorbers; c) the stack of sixteen spectra of star--forming
  galaxies at $\langle z\rangle=1.61$ that belong to the overdensity; d) the
  stacked spectrum of $\langle z\rangle~0$ starburst galaxies by Leitherer
  et~al. (2010).
\label{fig:stack_lbgabs}}
\end{figure}

\section{Large Amounts Of $T\sim~10^4$ K Gas In The Overdensity At $z\approx 1.61$}

The three LBGs discussed above share the common property of having absorption
features, all located approximately at the same wavelength, $\lambda\sim 7300$
\AA, that do not match any known strong spectral feature at their
redshifts. These ``spurious'' lines all match, however, the rest--frame
wavelength of \ion{Mg}{2}~$\lambda2800$ absorption at the redshift of the
intervening galaxy overdensity at $z\approx 1.61$. Since no other feature is
observed in the spectrum of any of the galaxies that would confirm this
interpretation, we have obtained a stacked spectrum by averaging each
candidate \ion{Mg}{2}\ absorption line after shifting its central wavelength
to the common rest--frame wavelength $\lambda=2800$ \AA. We found that it
makes no difference in the result if the spectra are scaled or not before the
average is computed, which is reasonable since the galaxies have similar
continuum intensity. Thus, the spectrum of galaxy G2 enters twice in the
stack, albeit each time with a different wavelength shift, since there are two
candidate \ion{Mg}{2}\ absorption lines in it. Such stacked spectrum is shown
in Figure \ref{fig:stack_lbgabs} (top spectrum).

The stacked absorption feature has rest--frame equivalent width $W_r=6.2\pm
1.5$ \AA\ and appears barely resolved into a doublet. We have fit the feature
to a two--component Gaussian line profile using a chi--square procedure to
measure the central wavelength of each component, finding best--fit values
$\lambda_G=2796.5\pm 0.7$ \AA\ for the blue one and $\lambda_G=2803.0\pm 1$
\AA\ for the red, respectively (we left the central wavelength and line width
of both components as free parameters finding the reduced chi--square of the
best fit to be $\chi_r^2\approx1.0$). Each component has rest equivalent width
$\sim 3$ \AA, and their width, ${\rm FWHM}\approx 4$ \AA, is consistent with
kinematic broadening comparable to the spectral resolution, i.e. $200$
\kms\ in the rest--frame. No \ion{Mg}{1}\ absorption line is observed in the
stacked spectrum, although none is be expected given the S/N if the ratio of
the equivalent width of this line to that of \ion{Mg}{2}\ is similar to what
observed the galaxy spectra discussed below, i.e. $\approx 1/6$.

\begin{figure}
\epsscale{0.9}
\plotone{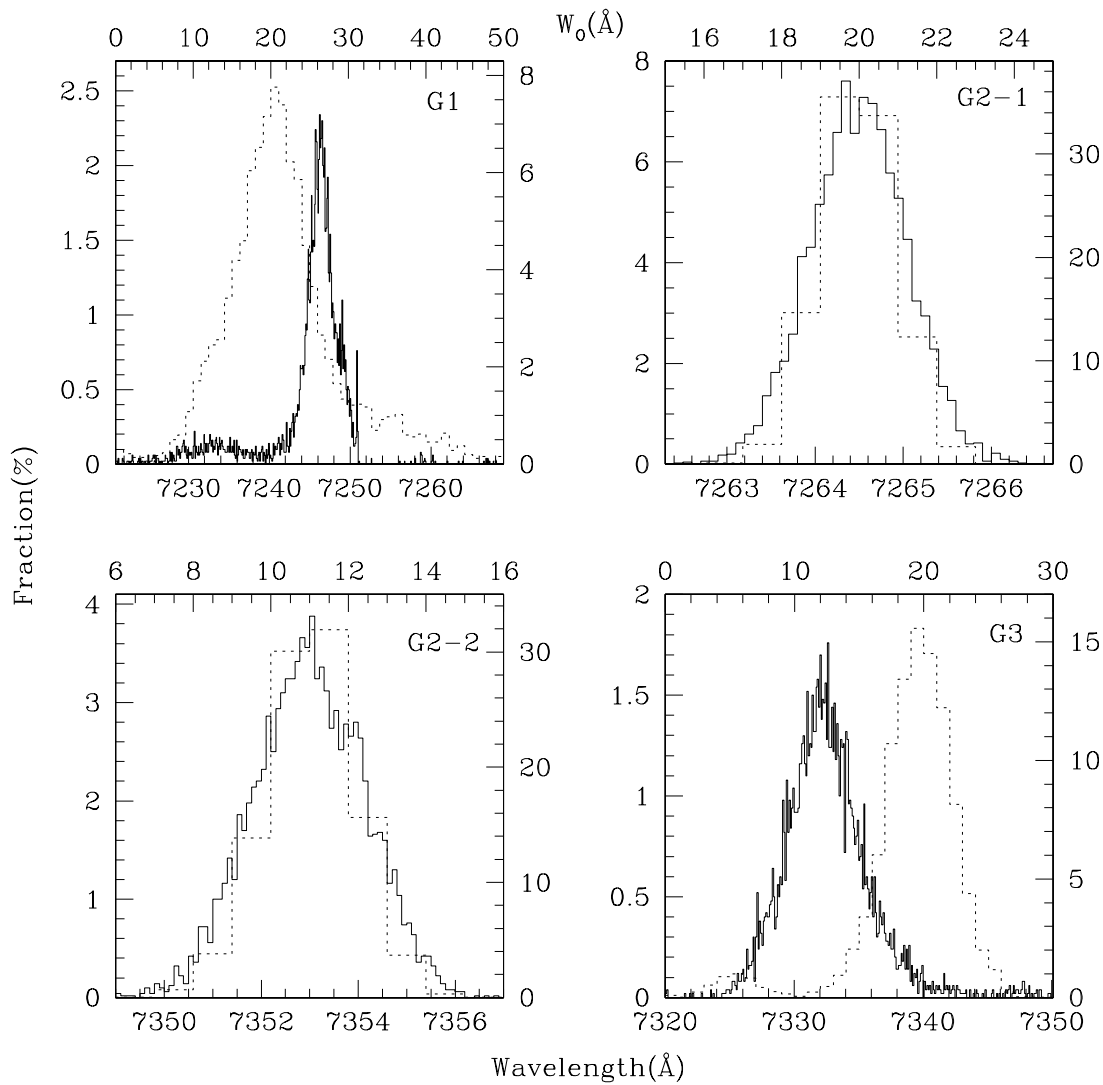}
\caption{The distribution of the wavelength of the centroid of the line (solid
  histogram, bottom abscissa) and of the equivalent width (dotted histogram,
  top abscissa) of 5,000 random realizations of the four intervening
  absorption features in the spectra of the three LBGs G1, G2 and G3 (G2 has
  two absorption features). The random realizations are obtained by varying
  the pixels in the spectral range $2705\le\lambda\le 2895$ \AA\ of each
  spectrum by a gaussian variate with null mean and variance equal to that
  spectrum's error array. 
\label{fig:lc_ew_hist}}
\end{figure}

\begin{figure}
\epsscale{0.9}
\plotone{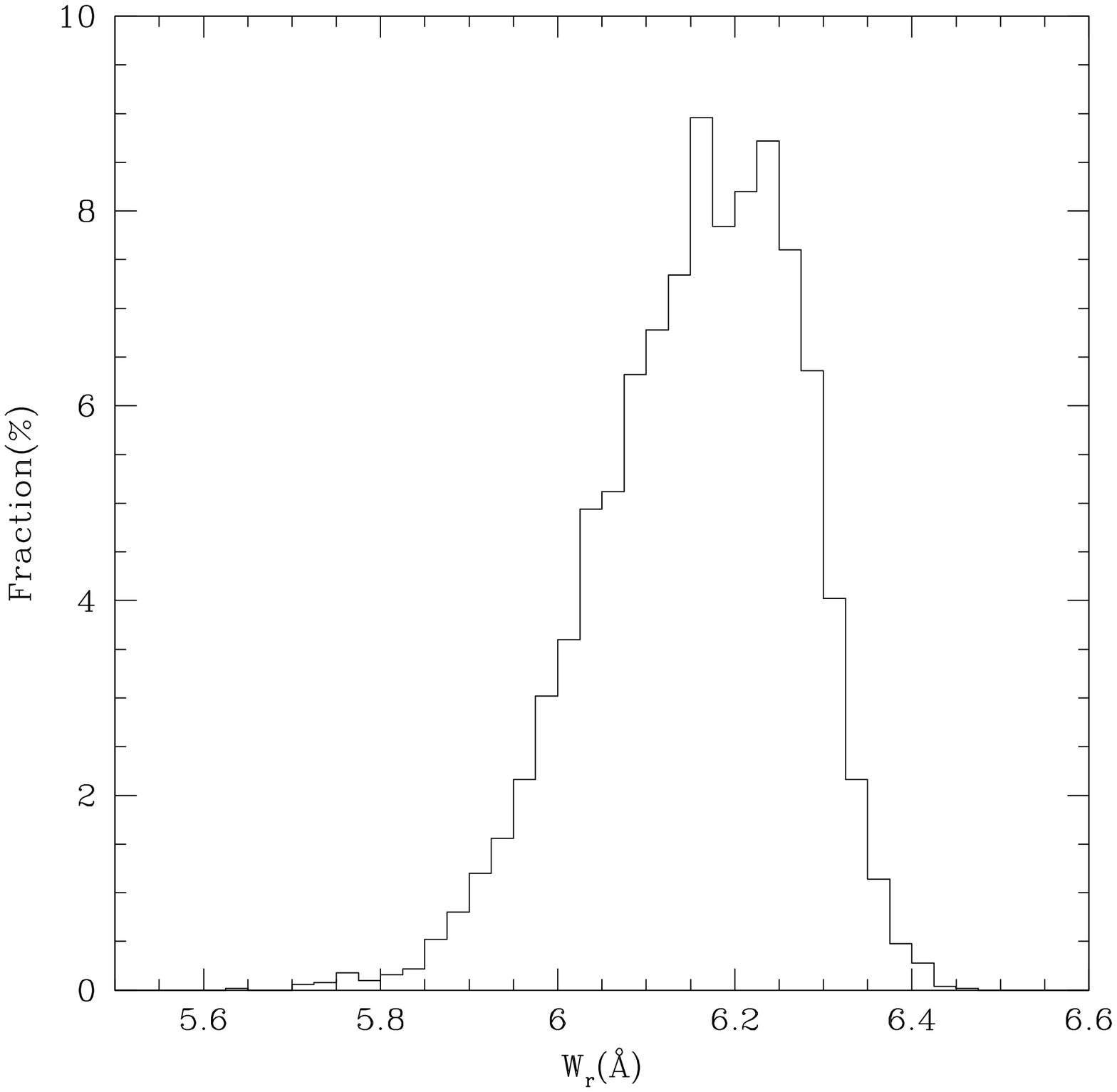}
\caption{The distribution of the equivalent width of the stack of the 5,000
  random realizations of the four intervening absorption features described in
  Figure \ref{fig:lc_ew_hist} (see also text).
\label{fig:stack_ew_hist}}
\end{figure}

Since the individual intervening absorption lines in the LBG spectra are
detected at relatively low significance level, in the range $3<\sigma<5$, we
have further investigated their significance using Monte Carlo
simulations. Specifically, we have created a large number of random
realizations (5,000) of the three LBG spectra by altering each spectrum, pixel
by pixel, with a gaussian variate with zero mean and variance from that
spectrum's error array. To include the uncertainty of the esitmate of the
value of the continuum in the simulations we have used all pixels in the
spectral range $2075\le \lambda\le 2895$ \AA, which is were the continuum is
measured with a linear fit (excluding the absorption line). We have then
fitted a gaussian profile to each intervening absorption feature, leaving the
centroid as a free parameter. For each random realization of the three spectra
we have also obtained the stack of the four absorption features, following the
same procedure adopted for the original spectra. The distributions of the
wavelength of the centroid and of the equivalent width of each absorption line
in the 5,000 realizations of the three LBG spectra are plotted in Figure
\ref{fig:lc_ew_hist}, which shows that in essentially all cases the lines are
reproduced with the same centroid wavelength and equivalent width as those
measured in the real spectra, within the $1-\sigma$ scatter. In the case of
the absorption line in G1's spectrum, the centroid of a fraction of the
realizations is found at significantly lower wavelength than that of the
majority of the realizations, at $\approx 7235$ \AA, consistent with the lower
S/N ratio of this feature compared to the other three. In all the other cases
the distributions of the centroid wavelength are well defined bell--shaped
curves. The distributions of the equivalent width of each of the four
absorption lines from the simulations, as well as that of their stack, the
latter reproduced in Figure \ref{fig:stack_ew_hist}, are also well defined
bell--shaped curves with width and peak value consistent with the
observations.  

Finally, we also used to simulations to assess the robustness of the detection
of a doublet in the stack of the four absorption lines. In general, a
two--component gaussian line profile provides a similarly good or better fit
to the simulated data than the single gaussian one, but this is to be expected
given the low S/N of the spectra and the larger number of free parameters of
the two--component gaussian profile relative to the single one. If we count
how often the double--gaussian model is consistent with the
\ion{Mg}{2}\ doublet, however, we find that only 562 of the 5,000 ($\approx
11$\%) realizations have flux and width ratio of the two components within
50\% from each other and with centroid separation in the range
$4\le\Delta\lambda\le8$ \AA\ around 2800 \AA, too few to suggest the doublet
is real, but too many to rule out its existence. In conclusion, the Monte
Carlo simulations show that the detection of the four intervening absorption
lines in the three LBG spectra is robust, but remain inconclusive on the
nature the marginally resolved doublet profile of the stacked absorption
line. In other words, given the quality of the data the Monte Carlo
simulations cannot confirm or recject the fact the the stacked absorption line
is marginally resolved into a doublet, as the data, taken at face vakue,
suggest. We note, however, that this feature has a very close morphological
similarity with the \ion{Mg}{2}\ absorption doublet in the co--added spectrum
of the seven GOODS galaxies at $\langle z\rangle\sim 1.9$ (spectrum b) of
Figure \ref{fig:stack_lbgabs}), which is also marginally resolved\footnote{The
  identification of this feature as the \ion{Mg}{2}\ doublet seems beyond
  question. The \ion{Fe}{2}\ $\lambda 2344,\lambda2382$ absorption lines and
  the \ion{Fe}{2}$^*$ $\lambda 2396$ emission line are clearly detected in the
  stacked spectrum. The \ion{Fe}{2}\ $\lambda 2587,\lambda 2600$ and
  \ion{Mg}{1} $\lambda 2852$ absorption lines are marginally detected if
  considered individually, but the fact all these these features are
  simultaneously observed as absorption features at the correct wavelength
  provides further evidence that that the identification of the spectral
  features is correct.}. The two co--added spectra have comparable S/N
characteristics , S/N$\approx 8$ and $\approx 12$ per resolution element,
respectively, thus suggesting that the doublet profile of the stack of the
four LBG absorption lines is real. Clearly, spectra with higher S/N are
necessary to solve this issue.

Figure \ref{fig:stack_lbgabs} also compares the co--added spectrum of the four
\ion{Mg}{2}\ absorbers to those of star--forming galaxies in the $z\sim 1.61$
overdensity as well as at $z\sim 0$, where the \ion{Mg}{2}\ and other
absorption features, such as the \ion{Fe}{2}, are produced in the galaxies'
ISM, including outflows, and the gas is illuminated by the UV light from
massive stars located in the galaxies themselves. As in the case of the LBG,
we have found that the stacked spectra are insensitive to scaling the
continuum of the individual spectra prior to calculating their average. The
most striking difference between the stacked spectrum of the LBG and those of
the galaxies is the systematic presence of \ion{Fe}{2}\ absorption in the
latter, while none is detected in the former. Finally, we point out that the
\ion{Mg}{2}\ and \ion{Fe}{2}\ absorption lines in the stacked spectrum of the
16 GOODS overdensity galaxies (spectrum c) in Figure \ref{fig:stack_lbgabs})
are clearly systematically blueshifted by $\approx 250$ \kms\ relative to the
systemic redshift, which is traced by the nebular emission lines (in all cases
the individual galaxy spectra are placed in the rest frame using the redshift
of the [\ion{O}{2}]~$\lambda3727$ emission line). This is evidence that the
gas participates in outflows, very likely powered by the activity of star
formation. The blueshifting signature of the outflows is not observed in the
$z\sim 1.9$ and $z\sim 0$ co--added spectra, because in this case the redshift
of the individual galaxies are measured from the \ion{Mg}{2}\ and
\ion{Fe}{2}\ absorption features themselves, since the [\ion{O}{2}] line is
outside of the available spectral range.

There is a hint that the \ion{Mg}{2}\ absorption in the stacked spectrum of
the $z\approx 1.61$ overdensity galaxies also includes a broader, asymmetric
component that extends {\it to the red of the doublet}, with maybe even
additional narrow discrete components (marked by short vertical segments)
over--imposed onto it. If real, this would be the signature of gas located in
front of the galaxies and moving toward them, since the putative absorbing gas
cloud is back--illuminated by the galaxies starlight and its absorption
feature is redshifted relative to the primary one. A similar kinematical
structure is observed in neither of the two Fe absorption features, which are
both detected with relatively higher S/N and whose wavelength is located in a
spectral region where the night sky emission is significantly fainter. It is
not observed in the stacked spectrum of the four LBG absorption systems, whose
S/N is lower and which are observed in the same spectral region as that of the
$z\approx 1.61$ galaxies, thus subject to the same night sky emission. In view
of the implications that such evidence, if confirmed, would have, we will
discuss it with more detail in the next section.

\begin{figure}
\epsscale{0.85}
\plotone{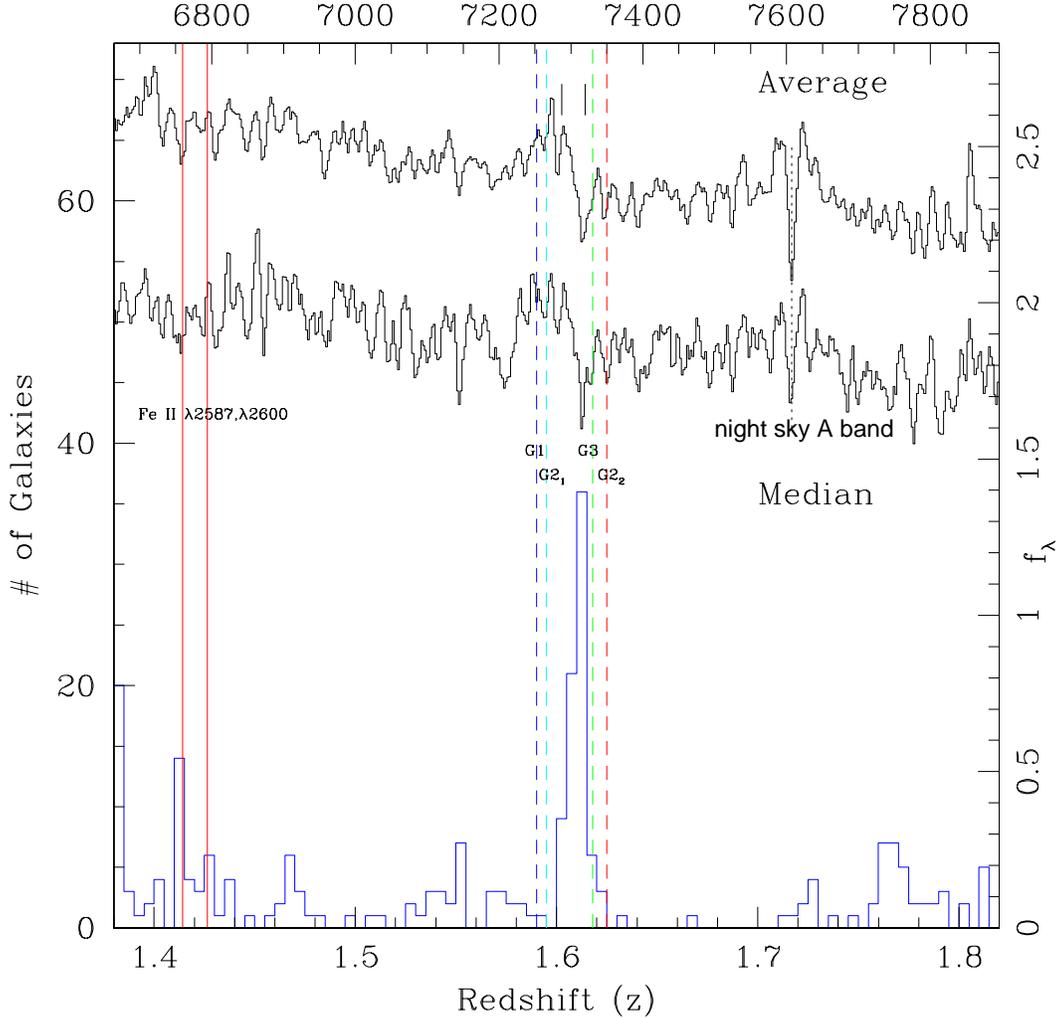}
\caption{The co--added spectrum of the 92 GMASS galaxies at $z>1.65$ in the
  background of the $z\approx 1.61$ overdensity, plotted together with the
  redshift histogram of the galaxies in the field (average top; median bottom;
  wavelength in \AA\ on the top horizontal axis; flux density in units of
  $10^{-19}\times $ erg s$^{-1}$ cm$^{-2}$ \AA$^{-1}$ on the right vertical
  axis; equivalent exposure time $T_{exp}=2070.5$ hr). The peak redshift of
  the overdensity coincides with a strong absorption feature (right short
  vertical segment) at $\lambda_c=7318\pm 1$ \AA, namely
  \ion{Mg}{2}~$\lambda2800$ at $z=1.6136$. Another weaker feature is possibly
  observed at $\lambda_c=7285$ \AA\ (left short vertical segment), at $\Delta
  V=-1351$ \kms\ from the first. Absorption features might also be present at
  the wavelengths of the two absorption systems of G2 (the four LBG absorbers
  are marked by vertical dashed lines). Two solid vertical lines mark
  \ion{Fe}{2}~$\lambda2587,\lambda2600$ at $z=1.6136$; no absorption with
  strength comparable to that of \ion{Mg}{2}\ is observed at these
  wavelengths, although one possible line is observed at $\Delta V\approx
  -190$ \kms\ from \ion{Fe}{2}~$\lambda2587$ and another one at $\Delta
  V\approx +430$ \kms\ from \ion{Fe}{2}~$\lambda2600$. These lines, however,
  coincides with \ion{Mg}{2}\ at the redshift of two spikes around $z\approx
  1.4$, suggesting that they are associated with these two
  overdensities. There appears to be a general correspondence between redshift
  spikes and \ion{Mg}{2}\ absorption features in the co--added spectrum.
\label{fig:gmass_back_stack}}
\end{figure}

We have further investigated the presence of ``intra--overdensity'' gas using
the deeper GMASS spectra. Specifically, we have co--added together the spectra
of galaxies with redshift $z>1.65$, namely in the background of the $z\approx
1.61$ overdensity but distributed along the same line of sight. We have
selected a total of 92 ``background'' galaxies from the GMASS survey, which
have apparent magnitude $z_{850}\le 25.8$ mag, and have visually inspected
each one of their spectra for the presence of intervening absorption lines,
finding none. Based on the S/N of the spectra, however, we estimate that we
would have identified absorption features comparable to the ones we found in
the GOODS LBG spectra (if they were there) only in galaxies with
$z_{850}\lesssim 25$ mag, namely 84 out of 92 (the GMASS spectra are
significantly deeper than the GOODS ones). We have subsequently co--added the
spectra of the 92 galaxies in the observed frame with no registration to any
common wavelength, with the the idea that if there is a \ion{Mg}{2}\ trough at
$z\approx 1.61$ in correspondence with the overdensity, this should give rise
to an absorption feature in the combined spectrum approximately at the
wavelength $\lambda\approx 2.61\times 2800$ \AA, even if absorption features
in the spectra of the individual galaxies might not have sufficient S/N for
detection. Figure \ref{fig:gmass_back_stack} shows the stacked spectrum
(average top; median bottom), which has the equivalent of $2,070.5$ hours of
integration time, plotted together with the redshift histogram of galaxies
(GOODS and GMASS) in the field. We have obtained stacks with and without
scaling each individual spectrum and found the results to be insensitive to
the weighting scheme.

The figure shows that the peak redshift of the $z\approx 1.61$ overdensity
coincides with a well--detected absorption feature in the co--added spectrum,
marked by a short vertical segment (right segment). The feature is present in
both the mean spectrum and in the median one, suggesting that it is not due to
a few galaxies but rather is representative of the average. The area of the
absorption feature and its equivalent width are $\sim 10$ times larger than
those of the weakest features in its proximity (regardless of whether these
weak features are noise or signal in this very deep spectrum), while its
width, ${\rm FWHM}\approx 23$ \AA, or $\approx 950$ \kms, is $\approx 5$ times
larger.  The central wavelength of the feature is $\lambda_c=7318\pm 1$ \AA,
corresponding to \ion{Mg}{2}~$\lambda2800$ observed at $z=1.6136$. Another
weaker feature also seems detected at $\lambda_c=7285.33$ \AA\ (left short
vertical segment), at velocity separation $\Delta V=-1351$ \kms\ from the
first. The central wavelengths of the four \ion{Mg}{2}\ absorption features in
the LBG spectra are marked with vertical dashed lines and labeled with the
name of the galaxies. Additional weak absorption features are observed in the
stacked spectrum at the wavelengths expected for \ion{Mg}{2} at the redshifts
of two of them, namely G2$_1$ at $z=1.595$ and G2$_2$ at $z=1.625$, both
observed in the spectrum of G2. Since none of these absorption features are
detected in the individual spectra of the background galaxies, it is difficult
to constrain the strength of the absorption systems from the equivalent width
in the stacked spectrum, because we do not know how many galaxies contribute
to the absorption and how much.

The displacement of these possible discrete absorption features by $\approx\pm
10^3$ \kms\ from the strong absorption feature in correspondence of the peak
of the overdensity redshift distribution, bears a striking similarity with the
velocity difference of four LBG \ion{Mg}{2}\ absorption systems from the same
redshift. We wonder if the fact that we have not found strong
\ion{Mg}{2}\ absorption in individual spectra at the peak redshift of the
overdensity is indication of some physical mechanism that reduces the covering
factor of dense gas clouds inside the overdensity itself. While we cannot
address this issue here, we note that the broad absorption feature of the
GMASS co--added spectrum at the overdensity redshift peak would tend to
prevent detection of weak, discrete absorptions at that redshift.

In any case, the existence of discrete absorption lines, including those in
the LBG spectra, together with the broad one in the co--added spectrum would
imply that the spatial distribution of the gas is patchy, with denser clouds
embedded in a more diffuse distribution, and consistent with a relatively
smaller covering factor for the high--column density gas (e.g. atomic hydrogen
column density $N_H\gtrsim 10^{19.2}$ cm$^{-2}$, see below). The velocity
width of the $z\approx 1.61$ \ion{Mg}{2}\ main absorption feature,
$\sigma\approx{\rm FWHM}/2.35=400$ \kms, in the stacked spectrum is comparable
to the redshift width of the overdensity, $z_{rms}=0.0383$ or $\approx 440$
\kms, suggesting that the kinematical broadening due to the motion of the gas
is significantly less.  More generally, the equivalent width of strong,
saturated lines such as, very likely, the four LBG \ion{Mg}{2}\ absorption
systems and those in the co--added spectrum, has a very weak dependence on the
column density of the absorbing trough, while it gives an approximate
indication of the velocity spread along the LOS. In the low S/N spectra of the
four LBG \ion{Mg}{2}\ absorbers the line width shows that the absorbing gas
has a velocity spread that is, crudely, less than a few hundred \kms\ along
the LOS, consistent with the kinematics inferred from the absorption feature
in the co--added spectrum.

Finally, we also note that a number of other absorption features in the
stacked spectrum are observed at the wavelengths of \ion{Mg}{2}\ absorption
lines at the redshifts of many of the peaks of the galaxy redshift
distribution. We will present further analysis of this composite spectrum (as
well as other ones) in a separate work.

On the left of Figure \ref{fig:gmass_back_stack}, two solid vertical lines
mark the expected position of the \ion{Fe}{2}~$\lambda2587$ and
\ion{Fe}{2}~$\lambda2600$ absorption lines at the same redshift of the
\ion{Mg}{2}\ absorption feature. There is no detected absorption in the
composite spectrum at the expected central wavelength of the Fe lines that is
as strong as the \ion{Mg}{2}\ absorption, suggesting that the contribution to
the latter by the IGM of individual galaxies is negligible. One weak feature
is observed at $\Delta V\approx -190$ \kms\ from the expected wavelength of
\ion{Fe}{2}~$\lambda2587$, but none is present at the wavelength of
\ion{Fe}{2}~$\lambda2600$, with the closest feature being observed at $\approx
+430$ \kms\ from it. We emphasize that the lack of detection of the
$\lambda2600$ feature is even more significant than that of the $\lambda2587$
one, since it is stronger (stronger oscillator strength). Finally, that fact
that two observed absorption features both coincide with spikes in the galaxy
redshift distribution at $z\approx 1.4$ suggests that they are related to
\ion{Mg}{2}\ gas associated with those overdensities, and not to
\ion{Fe}{2}\ gas associated with the $z\approx 1.61$ one.

In conclusion, the analysis of the GMASS composite spectrum of galaxies at
$z>1.65$, i.e. in the background of the overdensity, provides additional,
independent evidence of absorption by a \ion{Mg}{2}\ trough associated with
the $z\approx 1.61$ galaxy overdensity whose chemical enrichment in terms of
Fe and Mg relative abundance is qualitatively similar to that observed in the
four LBG absorption systems. There is also evidence that the typical velocity
fields of the gas within the overdensity is a few $10^2$ \kms\ and that the
spatial distribution of the gas is patchy and consistent with high--column
density clouds having a relatively small covering factor.

\begin{figure}
\epsscale{0.85}
\plotone{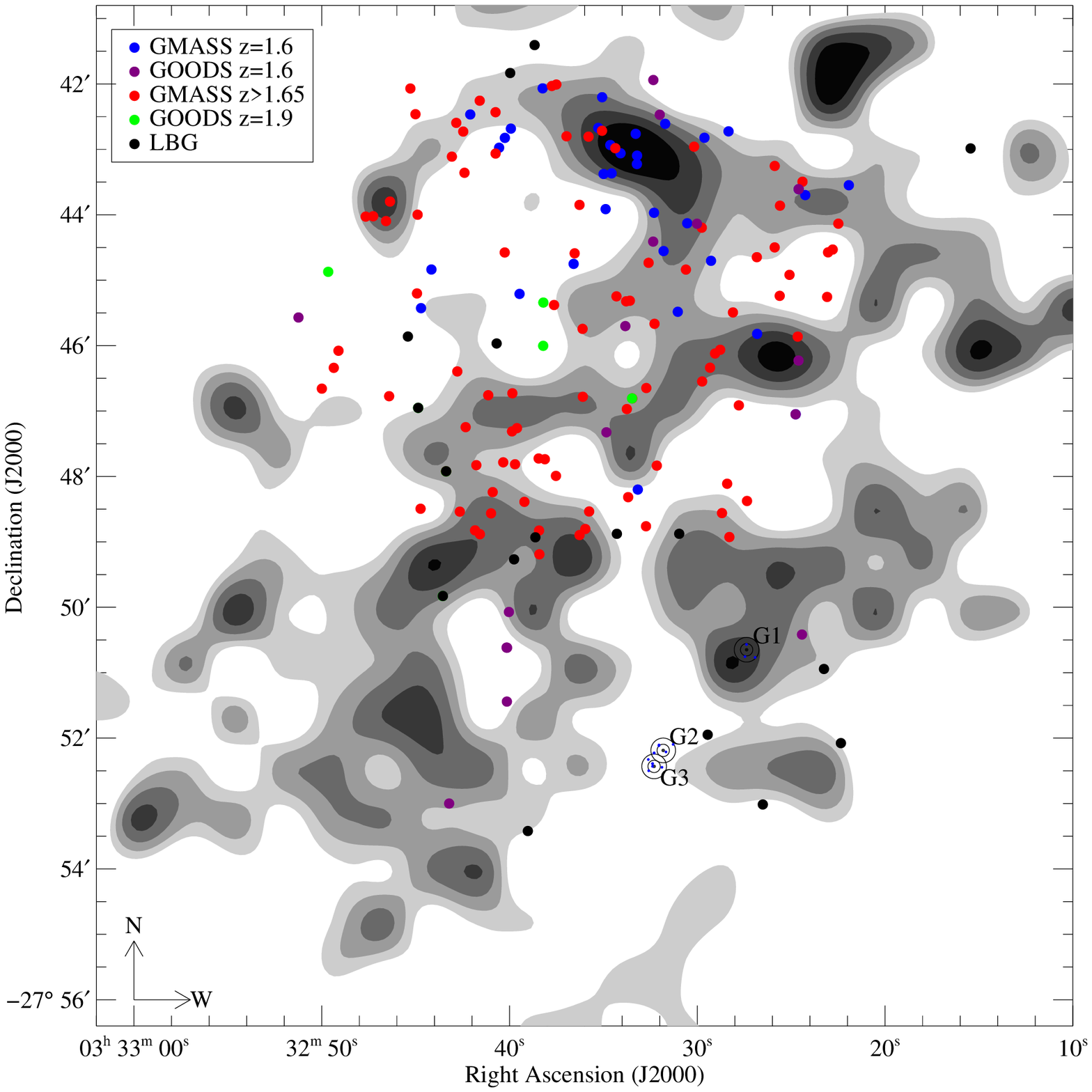}
\caption{The position on the plane of the sky of the galaxies with known
  redshift from the GOODS and GMASS spectroscopic surveys in the GOODS--South
  field discussed in the text. The three LBG galaxies with intervening
  \ion{Mg}{2}\ absorption systems are labeled. Also shown with gray scale
  contours is the surface density of galaxies with photometric redshifts from
  Dahlen et~al. (2010), in the range $1.56<z_{phot}<1.64$, i.e. candidate
  members of the $z\sim 1.6$ overdensity. The contours indicate 0.0, 0.25 0.5
  and $0.75\times $ the maximum overdensity above the median surface density
  of galaxies at $z\sim 1.61$.
\label{fig:overdens_dahlen}}
\end{figure}

\begin{figure}
\epsscale{0.85}
\plotone{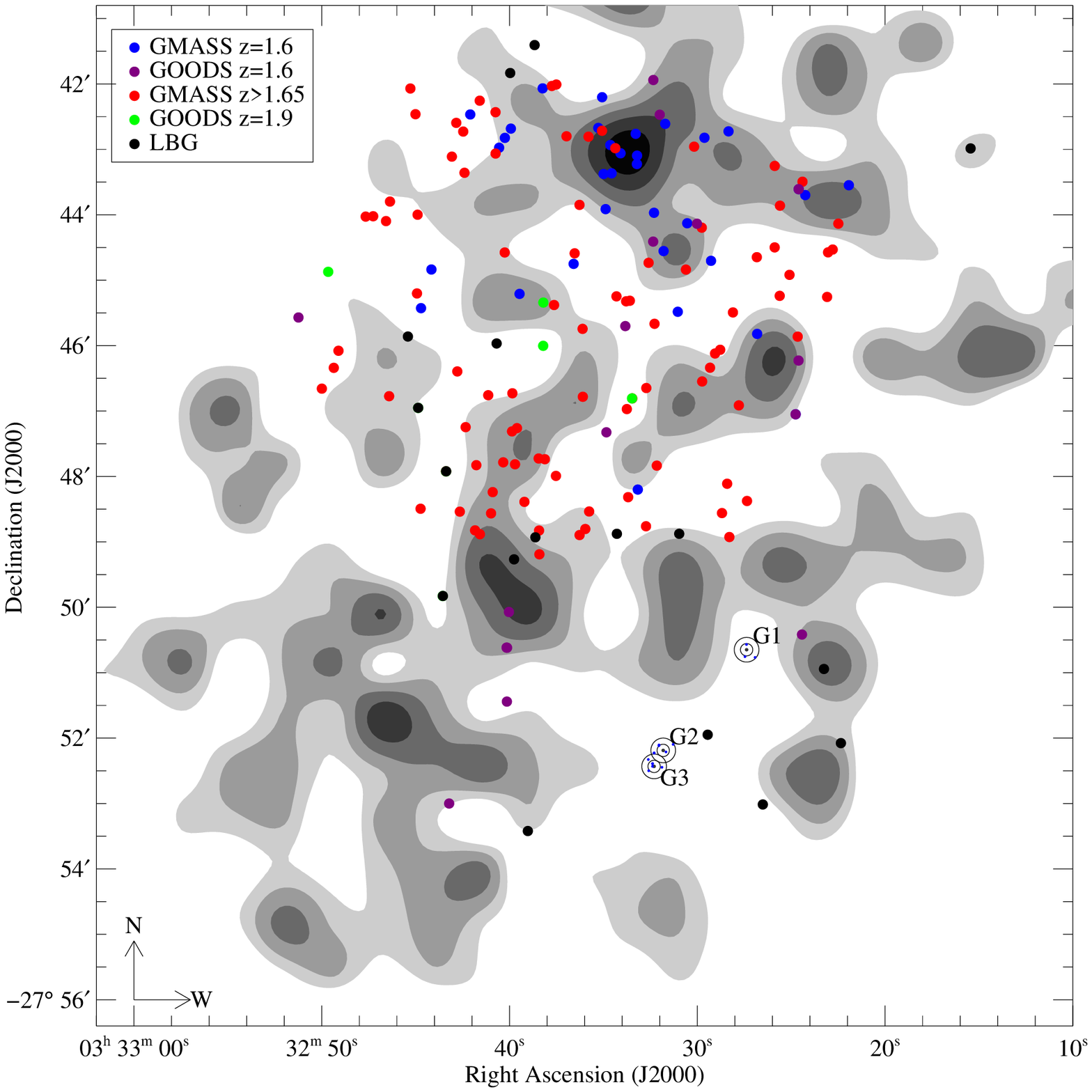}
\caption{As the previous Figure, but using the photometric redshift by Guo
  et~al. (2011, in preparation). 
\label{fig:overdens_guo}}
\end{figure}

Figures \ref{fig:overdens_dahlen} and \ref{fig:overdens_guo} show the position
on the plane of the sky of the galaxies from the GOODS and GMASS spectroscopic
samples, plotted together with surface density contours of galaxies with
photometric redshift in the range $1.56<z_{phot}<1.64$, i.e. candidate members
of the $z\sim 1.61$ overdensity. To assess the robustness of the photometric
redshifts is tracing large--scale structures, we used two independent photo--z
catalogs in the GOODS--South field, one made by Dahlen et~al. (2010) and the
other by Guo et~al. (2011, in preparation; see also Cassata et~al. 2010 for a
brief description of this catalog); both take advantage of the 12--band GOODS
panchromatic photometric catalog (UBVizJHK plus the four IRAC bands at 3.6,
4.5,5.8 and 8.0 $\mu$m). As the figures show, most of the large density peaks
are observed at the same position in both catalogs, although the morphology of
the regions with lower density contrast varies between the two catalogs. In
particular, galaxy G1 is in the background of a relatively over--dense region
in Figure \ref{fig:overdens_dahlen}, and of an under--dense one in Figure
\ref{fig:overdens_guo}. As noted before, galaxies G1, G2 and G3 are
approximately aligned along a line and are on the outside of the region where
most of the GMASS spectroscopically identified galaxies, either belonging to
the overdensity or in the background, are located. The spatial extension of
the overdensity, however, seems to be much larger than this region, covering
at least all of the extent of the GOODS field, approximately 10 arcmin by 16
arcmin, or 13.3 Mpc by 21.2 Mpc at $z=1.61$ (comoving), and perhaps more.

In summary, from intervening \ion{Mg}{2}~$\lambda\lambda2797,2803$
absorption features in the spectra of three LBG at $3<z<3.5$ and in the
average stacked spectrum of 92 galaxies at $z>1.65$ (not including the LBG) we
have identified gas apparently associated with a large overdensity of galaxies
at $z\approx 1.61$, whose spatial distribution extends at least over the same
projected linear size of the overdensity, namely $\gtrsim 10$ Mpc, and with a
likely patchy distribution of column density. From both the spectra of the LBG
and that of the 92 $z>1.65$ galaxies there is no evidence of corresponding
\ion{Fe}{2} absorption, a feature that has approximately equal strength in the
spectra of the outflows from star--forming galaxies at similar redshift,
including the galaxies in the overdensity, suggesting that the
``intra--overdensity'' gas is chemically more primitive than that of the
galaxies' ISM, as we shall discuss below.

\section{The Nature Of The Four LBG Absorption Systems}

\begin{figure}
\epsscale{1.0}
\plotone{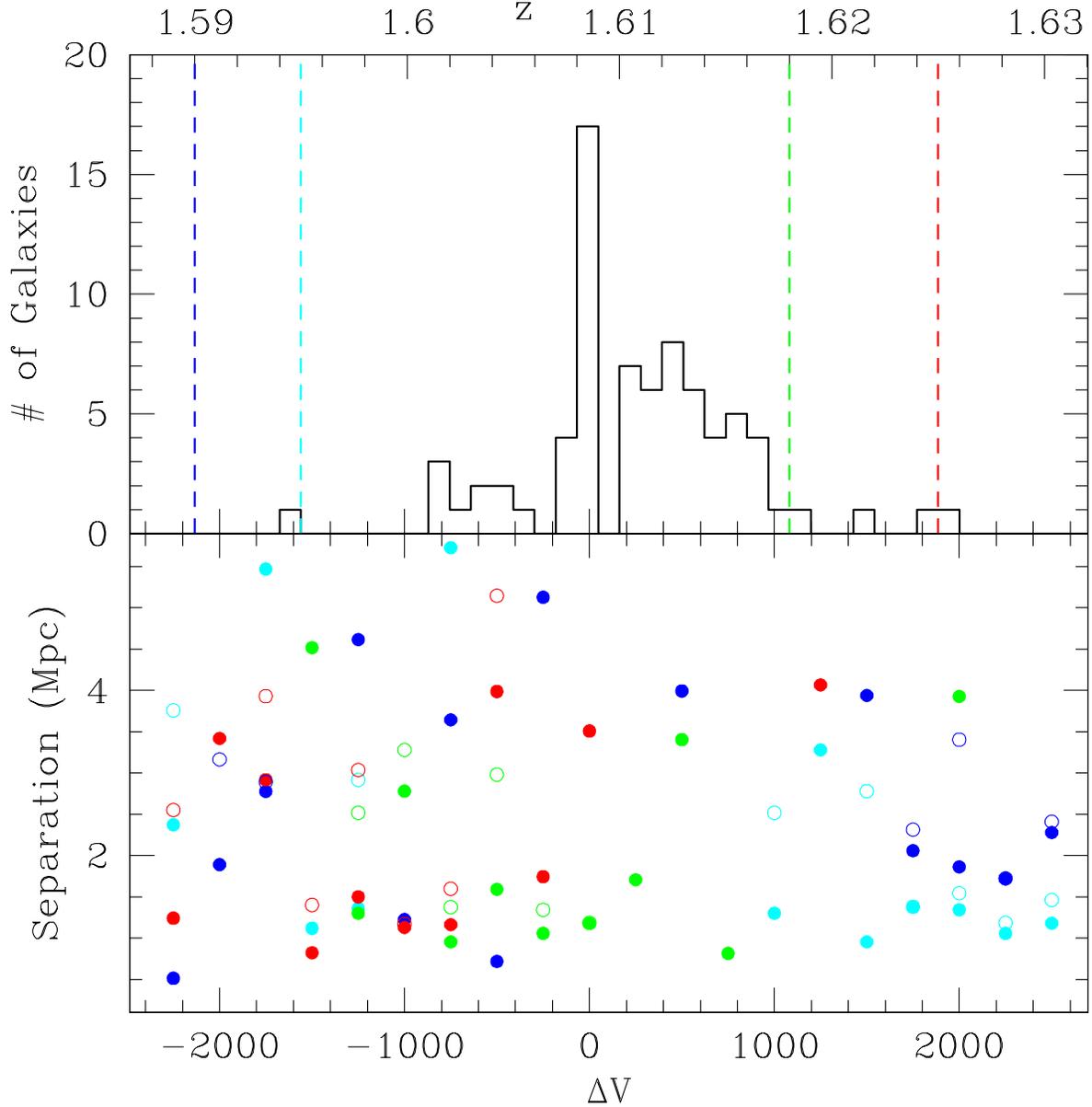}
\caption{{\bf Top.} Histogram of known spectroscopic redshifts (top scale) and
  velocity difference from the median redshift, $\langle z\rangle=1.609$
  (bottom scale), of galaxies in, or in proximity of, the $z\approx 1.6$
  overdensity, together with the redshift of the central wavelength of the
  four \ion{Mg}{2}\ absorbers. {\bf Bottom.} The projected distance at
  $z=1.609$ between the LOS from each \ion{Mg}{2}\ absorber, color--coded as
  the top figure, to the nearest (solid circles) and second nearest (open
  circles) galaxy with known spectroscopic redshift as a function of velocity
  separation. These galaxies have been selected from the GOODS/Spizer 4.5
  \micron\ images to satisfy the flux limit $m_{4.5\mu m}\le 23$ mag,
  essentially a mass--limit selection criterion approximately corresponding to
  $m_{star}<2\times 10^{9}$ \msun, with a small dependence on the spectral
  type. There is no evidence that the absorbers are associated with any known
  bright galaxy in the field.
\label{fig:dvhist}}
\end{figure}

The top of Figure \ref{fig:dvhist} shows the redshift of the four LBG
intervening \ion{Mg}{2}\ absorption systems as well as the redshift
distribution of galaxies in the field. It is clear that the four
\ion{Mg}{2}\ absorbers are all located in the tails of the
overdensity.\footnote{We note that the Kurk et~al. (2009) suggest the presence
  of sub--structure within the overdensity, with one smaller group of galaxies
  centered around redshift $z=1.59$ and a larger one at $z=1.61$,
  corresponding to a velocity separation $\Delta v\approx 2300$ \kms.} Two
factors likely contribute to the fact that the redshift of the absorbers
apparently ``avoid'' those of most galaxies in the overdensity. The first is a
selection effect introduced by the band of night--sky emission lines at
$7200\lesssim\lambda<\lesssim7400$ \AA, the spectral region where
\ion{Mg}{2}\ falls for most galaxies in the overdensity. These lines bias the
observations against the detection of absorption features in relatively faint
spectra, such as our sample of LBG (note that the absorber of G1 and the first
absorber of G2 are observed to the blue of the sky--line band, while the
second absorber of G2 and that of G3 are observed at the wavelength of weaker
lines of the band). The second factor is the relatively limited size of the
sample of background galaxies that are available to us, which include the 21
GOODS galaxies and the 84 GMASS ones that have adequate S/N for detection of
intervening absorption systems. Hints of the presence of discrete features to
the blue and to the red of the redshift distribution peak are also present in
the co-added spectrum of the GMASS galaxies in the background of the
overdensity shown in Figure \ref{fig:gmass_back_stack}, as we have observed in
the previous section. In that spectrum, no discrete features can be detected
in the proximity of the peak, however, because of the presence of broad
absorption lines. Thus, while it is possible that some physical reasons might
be responsible for the lack of discrete absorption features at small velocity
separation from the peak of the galaxy redshift distribution, our current data
do not allow us to address the issue at this time.

The bottom of Figure \ref{fig:dvhist} shows the shortest and second--shortest
impact parameter from the LOS of each absorber to that of the whole (GOODS and 
GMASS) spectroscopic sample as a function of the velocity separation. While
for any given \ion{Mg}{2}\ absorber there is at least one galaxy with velocity
separation of $\Delta V\approx 500$ \kms\ or less, the impact parameter in
each case is $d>400$ kpc. In fact, in the range $1.55<z<1.65$, where the
spectroscopic sample is on average $\approx50$\% complete over the $\approx
0.04$ square degree of the GOODS--South field down to $z_{850}\lesssim25$ mag
(the approximate limit if the spectroscopic observations), there are no
spectroscopically--identified galaxies (or galaxies that satisfy the selection
criteria for the spectroscopic observation, i.e. $m_{4.5\mu}\le 23$) with impact
parameter to the LOS of any of the absorbers (at $z=1.61$) shorter than
$d=400$ kpc.

\begin{figure}
\epsscale{1.0}
\plotone{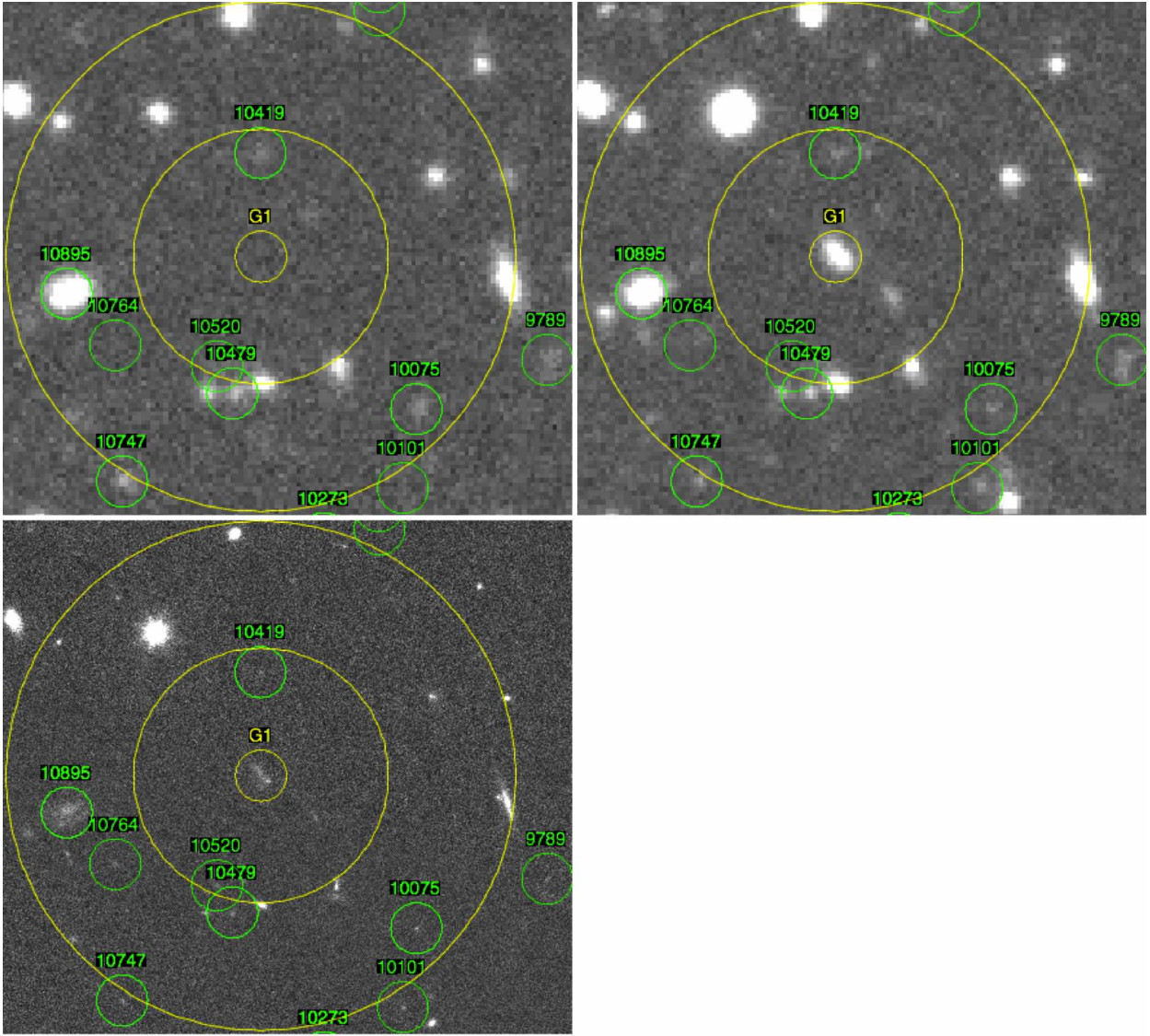}
\caption{Images in the U (top left), R (top right) and ACS \wz\ (bottom left)
  bands of the region around the sight line toward galaxy G1. The LBG is at
  the center of the concentric circles, whose radius corresponds to impact
  parameter of 10, 50 and 100 kpc at $z=1.61$. The LBG is not detected in the
  U band, since in the rest--frame UV spectrum of the galaxy the transmittance
  of the filter lies entirely blueward of the 912 \AA\ Lyman continuum
  break. Galaxies in the GOODS--South V2 catalogs whose photometric redshift, 
  measured by either the Dahlen et~al. (2010) or by Guo et~al. (2010), falls
  in the range $1.5<z_p<1.7$ are also marked by a circle and labeled with
  their catalog ID. The unmarked galaxy to the SW of the LBG roughyl mid way
  between the 10 and 50 kpc circles is very likely at the same redshift, as
  evidenced by the fact that it is undetected in the U band.
\label{fig:G1_img}}
\end{figure}

\begin{figure}
\epsscale{1.0}
\plotone{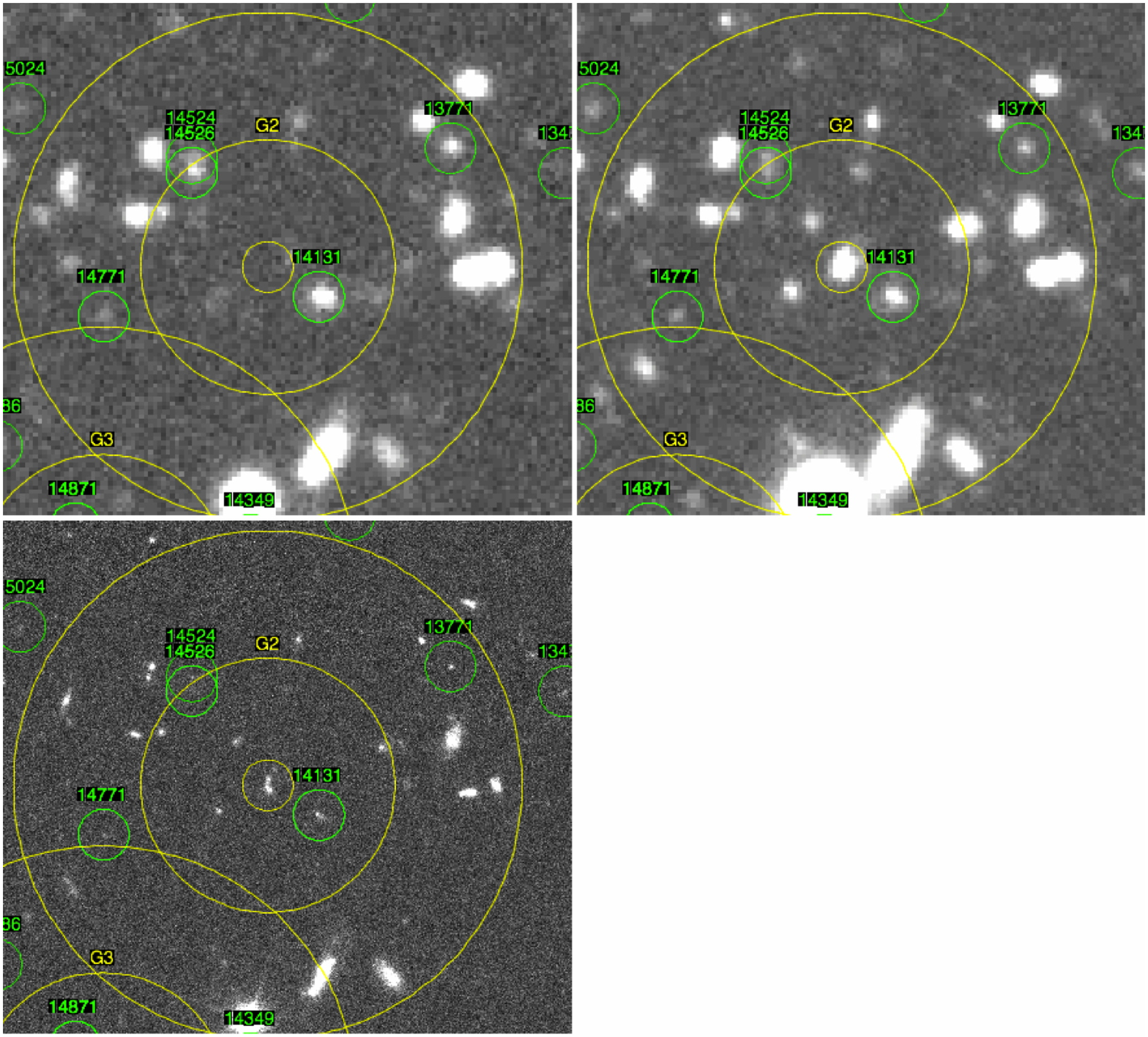}
\caption{As in Figure \ref{fig:G1_img}, but for galaxy G2. Thr three unmarked
  galaxies between the 10 and 50 kpc circles and the one on the the 50 kpc
  circle to the ENE of the LBG are very likely at redshift $z>3$ as shown by
  their colors (all are U--band dropouts). 
\label{fig:G2_img}}
\end{figure}

\begin{figure}
\epsscale{1.0}
\plotone{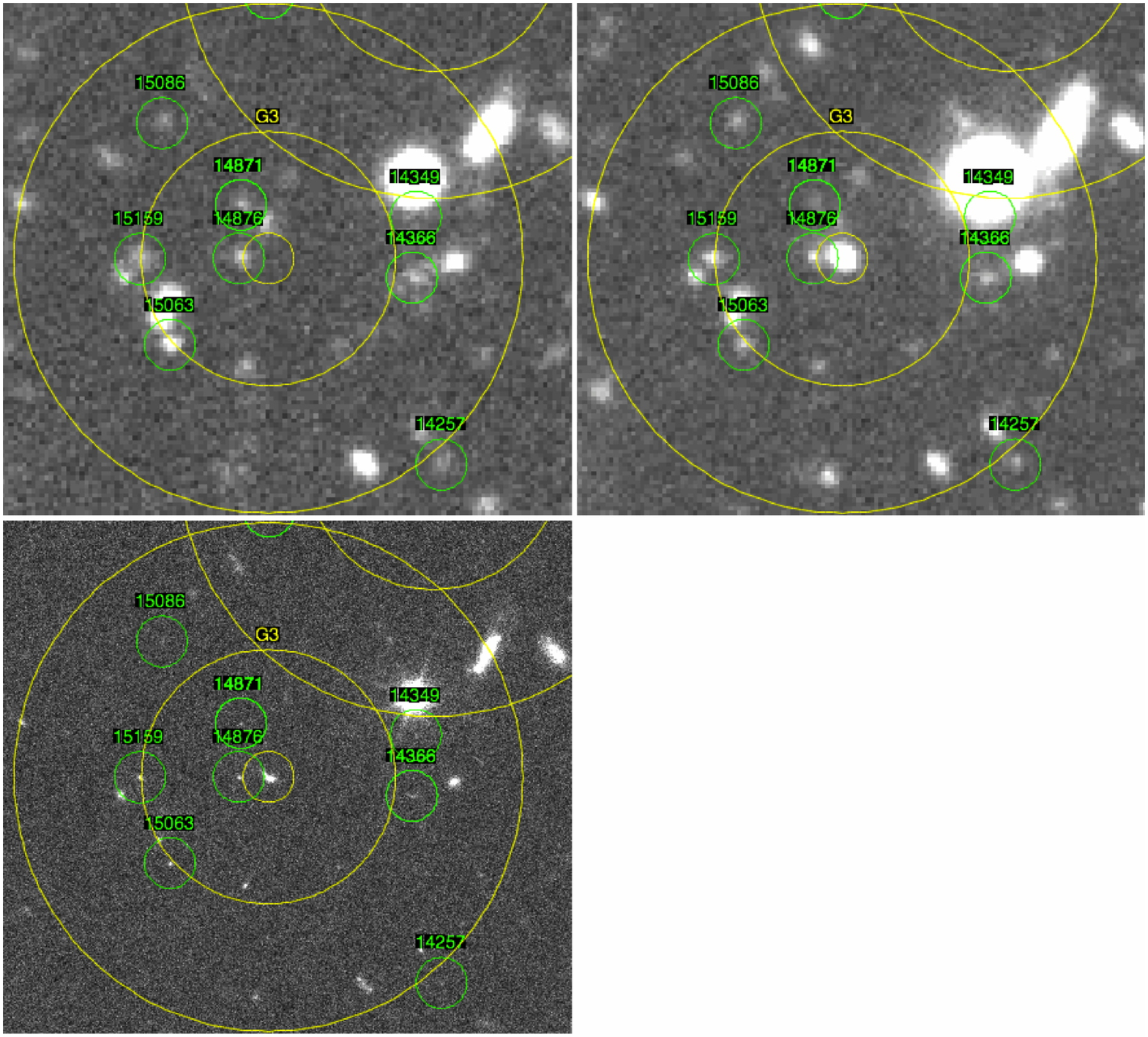}
\caption{As in Figure \ref{fig:G1_img} and \ref{fig:G2_img}, but for galaxy
  G3. All the unmarked galaxies inside the 50 kpc circle, even if they had
  redshift $z\sim 1.61$, are too faint to be responsible for the ultra--strong
  \ion{Mg}{2}\ absorption feature in the spectrum of G3 (see text).
\label{fig:G3_img}}
\end{figure}

We have also used photometric redshifts to look if galaxies either lacking
redshift identification or fainter than the spectroscopic sample's limiting
flux might be plausibly responsible for the observed \ion{Mg}{2}\ absorption
systems. Down to the flux limit $m_{4.5\mu m}<23$ mag of the spectroscopic
sample, no galaxy with photometric redshift in the range $1.5<z_p<1.7$ in
either of the two lists that we have considered has impact parameter shorter
than 220 kpc from any of the three LBG. We have looked for fainter galaxies
both in the ACS \wz\ image, as well as in the ultra--deep images in the U and
R band recently obtained by Nonino et al (2009) with VLT/VIMOS. The
sensitivity of the GOODS ACS images (v2.0) has been discussed elsewhere
(e.g. see Lee et~al. 2009 and Dahlen et~al. 2010); here we simply point out
that the depth of the images translates into stellar mass sensitivity (see
Table 1) significantly smaller than that of galaxies thought to be capable of
producing ultra--strong \ion{Mg}{2}\ absorption systems, as we shall discuss
below. Figures \ref{fig:G1_img}, \ref{fig:G2_img}, and \ref{fig:G3_img} show
the images of the regions around the three LBG. In each case the LBG is at the
center of the concentric circles, which have a radius of 10, 50 and 100 kpc at
$z=1.6$ (1.18, 5.9 and 11.8 arcsec, respectively).  Galaxies with photometric
redshift in the range $1.5<z_p<1.7$ in {\it either} the Dahlen's {\it or}
Guo's photo--z catalog are also shown at the center of 10--kpc circles with
their GOODS v2.0 ID labeled. Table 1 summarizes relevant properties of these
galaxies.

\begin{deluxetable}{llllllllll}
\tablecaption{Galaxies Around the LBGs' LOS\label{tbl-1}}
\tablewidth{0pt}
\tablehead{
\colhead{LBG} & \colhead{ID\tablenotemark{a}} & 
\colhead{$z_p$\tablenotemark{b}} & \colhead{$z_w$\tablenotemark{b}} & 
\colhead{$z_p$\tablenotemark{c}} & \colhead{$z_w$\tablenotemark{c}} & 
\colhead{\wz} & \colhead{$K_s$\tablenotemark{d}} & \colhead{$d$\tablenotemark{e}} & \colhead{$Log(M)$\tablenotemark{f}}
}
\startdata
G1 &       &      &       &      &       &       &       &      &     \\
   & 10520 & 2.16 & 1.976 & 1.60 & 1.434 & 27.13 &\nodata& 46.4 & 9.3 \\
   & 10479 & 1.59 & 1.583 & 0.01 & 0.265 & 26.48 & 27.82 & 54.8 & 6.4 \\
   & 10764 & 1.27 & 1.592 & 1.00 & 1.058 & 27.15 & 25.48 & 67.0 & 8.2 \\
   & 10075 & 1.58 & 1.525 & 0.01 & 1.153 & 26.87 & 26.37 & 85.5 & 7.4 \\
G2 &       &      &       &      &       &       &       &      &     \\
   & 14131 & 1.60 & 1.600 & 1.73 & 1.769 & 25.40 & 24.75 & 23.3 & 9.1 \\
   & 14526 & 1.91 & 1.911 & 1.57 & 1.531 & 26.58 & 25.71 & 47.6 & 8.4 \\
   & 14524 & 1.90 & 1.999 & 0.05 & 1.678 & 27.29 & 25.13 & 52.1 & 7.5 \\
   & 14771 & 1.60 & 1.800 & 1.84 & 1.501 & 27.29 & 25.87 & 67.4 & 8.2 \\
   & 13771 & 1.60 & 1.495 & 0.89 & 1.098 & 26.48 & 25.73 & 85.7 & 7.5 \\
G3 &       &      &       &      &       &       &       &      &     \\
   & 14876 & 1.69 & 1.746 & 1.84 & 1.762 & 26.03 & 25.02 & 11.6 & 9.0 \\
   & 14871 & 1.60 & 1.649 & 1.63 & 1.534 & 27.71 & 29.01 & 23.7 & 7.8 \\
   & 15159 & 1.73 & 1.797 & 1.70 & 1.725 & 25.64 & 24.12 & 50.5 & 9.3 \\
   & 15063 & 1.59 & 1.481 & 1.23 & 1.107 & 26.37 & 26.87 & 51.5 & 7.7 \\
   & 14366 & 1.11 & 1.246 & 0.93 & 0.876 & 26.08 & 24.44 & 56.9 & 8.2 \\
   & 14349 & 2.07 & 2.211 & 1.64 & 1.748 & 27.34 & 25.19 & 60.3 & 9.4 \\
   & 15086 & 1.60 & 1.996 & 0.62 & 0.606 & 26.90 & 24.67 & 67.7 & 7.1 \\
\enddata
\tablenotetext{a}{ID in the GOODS--South v2.0 ACS source catalog.}
\tablenotetext{b}{Peak and weighted photometric redshift from Dahlen et~al. 
  (2010).}
\tablenotetext{c}{Peak and weighted photometric redshift from Guo et~al. 
  (2010).}
\tablenotetext{d}{$K_s$--band magnitude, in the AB system (see Giavalisco
  et~al. 2004 and Dahlen et~al. (2010)).}
\tablenotetext{e}{Impact parameter to the sight line to the LBG at $z=1.6$,
  in kpc.}
\tablenotetext{f}{Stellar mass from SED fitting to spectral population
  synthesis model by Guo et~al. (2010).}
\end{deluxetable}

The three Lyman--break galaxies are not detected in the U--band images, since
these probe the SED beyond their Lyman limit (they have been selected as
B--band dropouts). This is very useful, because it gives us unrestricted
visibility at $z\approx 1.61$, down to the sensitivity limit of the image, to
check the possible presence of faint galaxies, undetected at redder
wavelengths because overlapping with the isophotal area of the LBG, which
could be responsible for the \ion{Mg}{2}\ absorption. We found that, in all
cases, no source is detected in the U band at the positions of the LBG down to
$\approx 30$ mag arcsec$^{-2}$ 1--$\sigma$ surface brightness limit (Nonino
et~al. 2009).

In the case of G1 there are no galaxies in the $1.5<z_p<1.7$ range with impact
parameter closer than 10 kpc, at the redshift of the \ion{Mg}{2}\ absorption
system, from the LOS to the LBG. The closest galaxy to G1, which has
photometric redshift $z\approx 3.5$ and is also undetected in the U (it is
barely detected in the \wz\ band, but clearly visible in the R one), is very
likely a close companion of the LBG itself. There is a very faint galaxy
within the 50 kpc radius, ${\rm ID}=10419$, with $z_{850}=27.20$ mag which, if
located at the redshift of the absorber, would correspond to rest--frame 3300
\AA\ absolute magnitude $M_{3300}=-17.1$. In the Guo et~al.'s catalog this
galaxy has peak and weighted photometric redshift $z_p=1.69$ and $z_w=1.4589$,
respectively, and stellar mass $Log(M)=8.4$.  Another two faint galaxies,
${\rm ID}=10520$ with $z_{850}=27.13$ mag and ${\rm ID}=10479$ with
$z_{850}=26.48$ mag, are located in close proximity of the 50 kpc radius; the
former barely enters in the selected photo--z window, while for the latter the
two measures of the photo--z are widely discrepant. As we shall discuss in the
next section, none of the galaxies around G1 is a good candidate for being the
host of the \ion{Mg}{2}\ absorbing trough.

Galaxy G2 has one galaxy within the 10 kpc circle and two galaxies inside the
50 kpc one, which all have high photometric redshifts ($z>3$), are undetected
in the U band and, thus, are likely spatially associated with it. The only
galaxy inside the 50 kpc circle with $1.5<z<1.7$, ${\rm ID}=14131$, has
$z_{850}=25.5$ mag and stellar mass $Log(M)=9.12$. As we will see below, its
mass seems too small by at least one order of magnitude to be able to host an
ultra--strong absorber, given the impact parameter. Also, even if this galaxy
were responsible for one of the two absorption systems of G2, it would be very
unlikely causing the other one, given the large velocity separation between
them. The fact that the second absorber would then remain without a galaxy
identification argues in favor of the interpretation that these LBG absorption
systems are not caused by gas associated with individual galaxies.

Galaxy G3 has two galaxies around it with relatively small impact parameter,
one just outside the 10 kpc circle (${\rm ID}=14876$) and the other at about
25 kpc (${\rm ID}=14871$). The former has $z_{850}=26.0$ mag, $K_s=25.0$ mag
and $Log(M)=9.0$, the latter $z_{850}=27.7$ mag, $K_s=29$ mag and
$Log(M)=7.8$. Again, as it will be clearer in the next section, both galaxies
seem too small and faint to qualify as likely candidate hosts of the absorbing
trough.

Finally, we note that while there is no strong bias against early--type
galaxies in the GMASS flux limit selection criterion, the spectroscopic
identification of these galaxies requires substantially higher sensitivity
than star--forming galaxies with the same $m_{4.5\mu m}$ apparent magnitude
because of the lack of emission lines in their spectra. As a result, it is
possible that, using only the spectroscopic sample, massive and
passively--evolving galaxies at $z\sim 1.61$ located in close proximity to the
LOS to the four LBGs, and thus potentially capable to host gas responsible for
the absorption systems, might have eluded identification. There is no evidence
of the presence of such galaxies around G1, G2 or G3 based on the deep URz
images discussed here.

\section{The Nature Of The Absorbing Gas}

The properties of the gas responsible for optically--thick \ion{Mg}{2}\ QSO
intervening absorption systems, and for the strong interstellar absorption
in the outflows of starburst galaxies, provide a very useful reference to
infer the nature of the gas responsible for our LBG \ion{Mg}{2}\ absorption
systems, as well as the \ion{Mg}{2}\ absorption observed in the co--added
GOODS and GMASS spectra.

\subsection{Gas Responsible For QSO \ion{Mg}{2}\ Intervening Absorption Systems}

A vast body of work on QSO absorption systems has characterized the properties
of the gas capable to give rise to strong intervening \ion{Mg}{2}\ absorption
features observed in their spectra. The main results that concerns this work
is that this low--ionization species is a tracer of a warm medium, $T\sim
10^4$ K, associated with atomic hydrogen, and that strong
\ion{Mg}{2}\ absorption systems, traditionally defined as those observed with
rest--frame equivalent width $W_r>0.3$ \AA, originate in optically--thick
\ion{H}{1}\ gas having column density $N_{HI}>10^{17}$ cm~$^{-2}$.  It must be
emphasized that this is a lower limit to the column density of atomic
hydrogen; it is very likely that the column density of the \ion{H}{1}\ gas
responsible for strong \ion{Mg}{2}\ absorption is much higher than this value
(see Steidel \& Sargent 1992). At relatively low redshift (e.g. $z\lesssim
1.5$) a key property of strong \ion{Mg}{2}\ absorption systems is that the gas
responsible for them is often found to be part of the extended gaseous halos
around massive, bright galaxies, roughly with $L/L^*>0.3$ and $M>3\times
10^{10}$ \msun, with little or no dependence on the spectral and morphological
type (e.g.  Steidel et~al. 1998, 2001; Churchill et~al. 2005).  Chen
et~al. (2010) find that the extent of the \ion{Mg}{2}\ absorbing gas around
galaxies depends strongly on their stellar mass $M_{star}$ and more weakly on
the specific star--formation rate SSFR, and suggest that this is naturally
explained if gas exchange to and from galaxies, i.e. outflows and inflows, are
the mechanisms that provide the gas responsible for the absorption.

The distribution of impact parameters, i.e. the distance from the line sight
to the QSO (the location of the absorbing clouds) to the center of the gas
host galaxies, for strong \ion{Mg}{2}\ absorption to take place has been found
to be in the range $15\lesssim d\lesssim 80$ kpc, with covering factor in the
range $0.5\lesssim C_F\lesssim 1$, depending on how the sight line to the QSO
is selected. For example, if \ion{Mg}{2}\ gas is known to be present at a
particular redshift because of the strong absorption system observed in a
QSO's spectrum, then there is a very high probability ($\sim 100$\%) that a
bright galaxy will be found nearby the LOS to the QSO as the likely host of
the gas (Steidel et~al. 1994). The converse case does not seem to be true.
Namely, the spectra of QSOs whose LOS has impact parameter to a known bright
galaxy in the above range have been observed to have strong intervening
\ion{Mg}{2}\ absorption systems at the redshift of the galaxy only roughly
$\sim 50$\% of times (Tripp \& Bowen 2005). Recently, Gauthier, Chen \& Tinker
(2009, 2010) and Bowen \& Chelouche (2010) found evidence that the covering
factor around the so--called Luminous Red Galaxies at $z\sim 0.5$, essentially
galaxies photometrically selected for having very low specific star--formation
rate (passively evolving), is significantly smaller, $C_F\sim 10$-15\%. They
suggest that either the halos of these galaxies are too hot to retain or
accreted cold gas, or these passively evolving galaxies simply do not have the
cool (i.e. $T\sim 10^4$ K) outflows to fill their halos with
\ion{Mg}{2}\ absorbing clouds.

Ultra--strong absorbers, i.e. $W_r>3$\AA, have also been observed in the
spectra of QSO, as well as star--forming galaxies where the gas is
back--illuminated by the galaxies' own UV light (Nestor et~al. 2005, 2007;
Tremonti et~al. 2007). They are between 2 and 3 orders of magnitude rarer
than weaker ones. These are generated by gas that resides either inside the
galaxies or in the outflows ejected from starburst ones. Nestor et~al.  find
that ultra--strong \ion{Mg}{2}\ absorption systems can also be caused by gas
in dwarf galaxies which are aligned along the LOS to the QSO, i.e. with the
LOS intersecting their optical light profile. In other words, to have
sufficient column density to give rise to to ultra--strong absorption, the gas
hosted by dwarf galaxies needs to be located at very small impact parameter
(less than a few kpc).

Nestor et~al. (2005) and Rao et~al. (2006) find evidence that a substantial
fraction of ultra--strong absorbers, perhaps $\gtrsim 65$\%, are also damped
Lyman--alpha (DLA) absorbers, implying that the \ion{H}{1}\ gas column density
exceeds the value $N_{HI}>2\times 10^{20}$ cm~$^{-2}$.  Conversely, every DLA
is also a \ion{Mg}{2}\ absorber (e.g. Rao \& Turnshek 2000). Although
\ion{Mg}{2}\ absorption can be generated by gas with column density up to
$\sim 3$ orders of magnitude smaller than that of DLA, e.g. $N_{HI}>10^{17}$
for $W_r>0.3$ \AA, at least the most metal--rich DLA have
\ion{Mg}{2}\ absorption with equivalent width in the range $2\lesssim
W_r\lesssim 3$ \AA\ (see Nestor et~al. 2007), and as many as $\approx 90$\% of
all DLA have $W_r\gtrsim 3$ \AA. Rubin et~al. (2010b) find that ultra--strong
\ion{Mg}{2}\ absorbers ($W_r>2.72$ \AA\ in their sample) produced in the
outflows from star--forming galaxies at $0.7<z<1.5$, which have stellar mass
and star-formation rate in the same range as the galaxies in our GOODS and
GMASS samples (see their Figure 8 and Kurk et~al. 2009), have atomic hydrogen
column density $N_H\gtrsim 10^{19.2}$ cm$^{-2}$. Note that the equivalent
width of these absorbers is smaller than the average equivalent width of our
four \ion{Mg}{2}\ LBG absorption systems (both components of the doublet),
namely $W_r=6.3\pm 1.5$ \AA; the smallest equivalent width among the four of
them, one of the two in G2's spectrum, has $W_r=5.4\pm 2$ \AA. It is also
smaller or comparable to the equivalent width of the individual components of
the doublet estimated above. 

Cases of optically--thick absorption systems (not necessarily
\ion{Mg}{2}\ systems) that are at least strong, i.e. $W_r>0.3$\AA, for which
an association with an individual bright galaxy could not be made are rather
rare. Examples include Tripp et~al. (2005), who reported of a sub--DLA
absorber with $N_H=10^{19.32}$ and remarkably low metallicity (${\rm
  [O/H]}=-1.60$) for which no galaxy could plausibly provide the gas to
explain the absorption (the nearest galaxy is a sub-$L^*$ with impact
parameter $d=86$ kpc, while the closest $L^*$ galaxy has $d=246$ kpc); and
Fumagalli et~al. (2010), who looked for the galaxies responsible for two QSO
DLA by imaging the area around the QSO at wavelengths below its Lyman limit, so
that the QSO's light is suppressed, offering unrestricted visibility along the
line of sight, and for one case they found no galaxy host.

In summary, the conclusion from all these works is that strong and
ultra--strong \ion{Mg}{2}\ absorption systems ($W_r>0.3$ \AA) require large
\ion{H}{1}\ column density, especially ultra--strong absorbers, whose column
density reaches the threshold value required for DLA. These conditions are
thought to be encountered in the gas flows that are being exchanged to and from
massive galaxies, either outflows, which are commonly observed in
star--forming galaxies at high redshift, or inflows, for which there has been
no direct evidence so far. In the case of the outflows, their large spatial
extent, physical conditions and geometry naturally explain many of the
observed properties of the absorbers, including the proximity of the line of
sight to a bright galaxy ($d<100$ kpc) and the inverse correlation between
equivalent width and impact parameters (see Steidel et~al. 2010). The case for
the inflow is much less clear. The gas is expected to come from large
separations from the galaxies, along what the theory predicts to be
filamentary streams. Thus, a high column density does not necessarily require
spatial proximity to a bright galaxy. The overall covering factor is expected
to be fairly small, however, revised to $\lesssim 3$\% around a $10^{12}$
\msun\ halo (Kimm et~al. 2010; A. Dekel, private communication) from early
estimates of $\sim 20$--25\% (Dekel et~al. 2009). This is consistent with the
paucity of inflow kinematic signature in most absorbers, as well as with the
rarity of absorbers with no galaxies responsible for them.

The equivalent width of our four ``LBG'' \ion{Mg}{2}\ systems qualifies them
as ultra--strong absorbers; however, differently from most ``QSO''
\ion{Mg}{2}\ absorption systems, intervening bright galaxies do not appear to
be directly responsible for them, as we have seen in Section 4. Also, note
that given the relatively low S/N ratio of our spectra, we cannot detect
absorption lines significantly weaker than ultra--strong. Thus, from the
spectra of our galaxies we do not have information on the presence and/or
location of gas capable of generating weaker absorption.

Finally, we conclude with a crude estimate of the mass of the \ion{H}{1}\ gas
associated with the absorbing trough observed in the LBG spectra, as well as
of the mass associated with the overdensity as a whole. Indeed, the resolved
nature of galaxies as probes of absorption systems lends itself to some
interesting inferences on the properties of the trough, such as a lower limit
to the mass of the gas. The large equivalent width of the four absorption
features and their minimum light intensity consistent with zero indicate that
the absorbing gas covers the full angular extent of the LBG. Hence, making an
educated assumption on the average \ion{H}{1}\ column density associated with
the \ion{Mg}{2}\ gas based on the known properties of similar
\ion{Mg}{2}\ absorbers that we have discussed, a lower limit to the mass of
the \ion{H}{1}\ gas clouds is
$$M_C\approx 2.5\times 10^8\times \Bigl({N_H\over 10^{20}~{\rm cm}^{-2}}\Bigr)
\times \Bigl({D\over 10 kpc}\Bigr)^2~\hbox{\msun},$$
where $D$ is the physical size at $z\approx 1.61$ subtended by the angular
diameter of a LBG at $z\sim 3$, i.e. the optically--bright (rest UV) part of
the galaxy that mostly contributes to the observed spectra, which is typically
5--7 kpc (Giavalisco et~al. 1996; Ferguson et~al. 2004; Ravindranath
et~al. 2006). The absorber's \ion{H}{1}\ mass, however, is almost certainly
larger, since it is very unlikely that it has the same angular extent of the
LBG.

In turn, a crude lower limit to the total mass of the gas can be derived from
the linear extent $R$ of the overdensity, a few Mpc, and the covering factor
of the high--column density gas, $C_F\approx 4/105=0.04$, namely the ratio of
number of observed absorbers (4) to the total number of spectra analyzed and
potentially of sufficient quality to reveal the presence of ultra--strong
absorption systems (21+84=105). We find
$$M_T > 10^{13}\times{C_F\over 0.04}\times{M_C\over 2.5\times
  10^8~\hbox{\msun}}\times\Bigl({R\over 2~\hbox{Mpc}}\Bigr)^2\times
\Bigl({D\over 10~\hbox{kpc}}\Bigr)^{-2}~\hbox{\msun}.$$

\subsection{The Metal Enrichment Of The $z\approx 1.61$ ``Intra--Overdensity'' Gas}

The lack of the \ion{Fe}{2}\ interstellar absorption lines, which generally
have strength similar to the \ion{Mg}{2}\ ones in star--forming galaxies (see
Leitherer et~al. 2010; Rubin et~al. 2010a,b; see also Nestor et~al. 2010 and
Tremonti et~al. 2007), in both the stacked spectrum of the four LBG absorption
systems (Figure \ref{fig:stack_lbgabs}, top) and in the GMASS stack of the
$z>1.65$ background galaxies (Figure \ref{fig:gmass_back_stack}), is the key
piece of evidence that the ``intra--overdensity'' gas is not directly
associated with, and is chemically older than, the ISM and/or the outflows of
galaxies in the overdensity.

In principle, the under--abundance of \ion{Fe}{2}\ could be due to depletion
of Fe by dust. While both iron and magnesium are prone to dust depletion, in
cold (e.g. $T<10^3$ K) and dense gas Fe can be significantly more depleted
than Mg (Savage \& Sembach 1996). The physical conditions for preferential
depletion of Fe over Mg could be encountered in the ISM in galactic
disks. However, the lack of massive, luminous galaxies ($L/L^*\gtrsim 0.3$ and
$M\gtrsim 10^{10}$ \msun) at $z\approx 1.61$ with small ($d\lesssim 100$ kpc)
impact parameter from the LOS to the absorbers argues against this
interpretation. The few galaxies whose impact parameter would be in the
plausible range to qualify as potential absorbers (assuming they have the
right redshift) are not sufficiently massive. In this situation it is more
likely that the absorption arises in the space between galaxies, at large
impact parameter ($d\gg 100$ kpc) from any bright one or around the
overdenisty as a whole. In both cases the gas is likely to have temperature
$T\sim 10^{4}$ K due to photoionization heating from the UV background and
density similar to that of other ultra--strong \ion{Mg}{2} absorbers, such as
the gas in the outflows (see Steidel et~al. 2010).  In these physical
conditions, the relative depletion of Mg and Fe should be similar. Moreover,
such regions (i.e., low-density outer gaseous halos of galaxies or
intergalactic gas clouds) generally contain little dust (e.g., Tripp et~al.
2002, 2005). A more likely explanation, therefore, is that the Fe
under--abundance reflects the nucleosynthetic history of the gas.  Magnesium
is an $\alpha$--element that is rapidly produced in Type II SNe, while Fe is
more slowly produced in Type Ia SNe, so Fe can be under--abundant compared to
Mg (e.g., McWilliam 1997).  This suggests that the absorbing gas is relatively
``chemically young,'' or at least younger than the gas in galactic outflows.
The signal-to-noise and spectral resolution of the stacked spectrum are not
adequate to support a detailed quantitative analysis, but the redshift is high
enough to bring a number of additional metal lines into the optical band, so
future follow-up studies will be valuable for constraining the nature of the
absorbing gas.

\begin{figure}
\epsscale{1.0}
\plotone{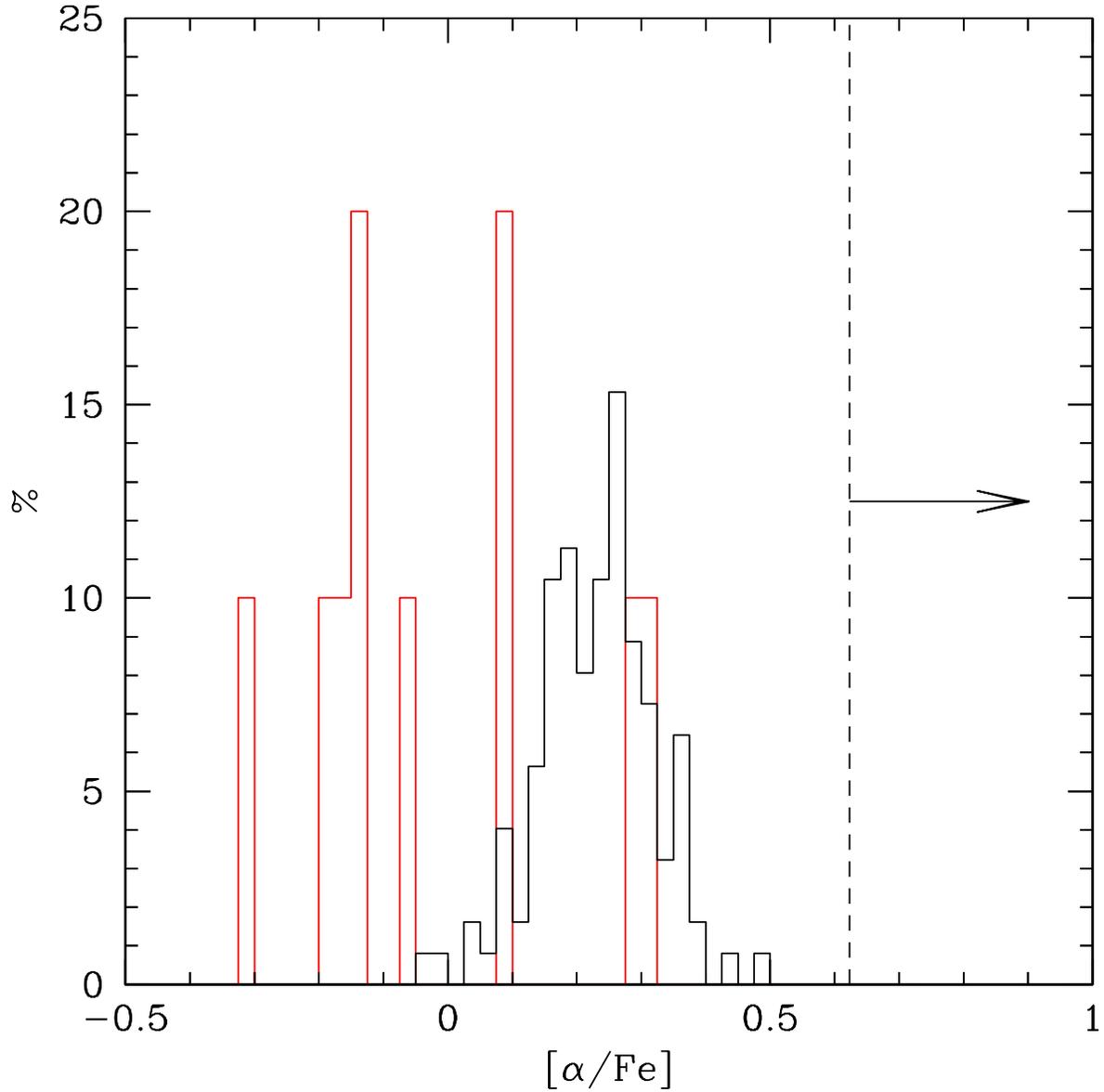}
\caption{The distribution of the ratio $\Big[\alpha\ Fe\Bigr]$ of
  $\alpha$--element to iron for early-type galaxies (E, S0 and cD) with
  velocity dispersion $27<\sigma <236$ \kms (red histogram, Sansom \&
  Northeast 2008) and $50<\sigma<360$ (black histogram, Thomas et~al. 2005)
  compared to the lower limit to the abundance ratio of Mg and Fe in our
  ``intra--overdensity'' gas (arrow departing from the vertical dashed line)
  estimated in the text.
\label{fig:Mg2Fe_abund}}
\end{figure}

To compare the chemical composition of our ``intra--overdensity'' gas to that
of the ISM of galaxies we need to translate the observed equivalent width
ratio into an abundance ratio. Unfortunately, this is not possible, because
the lines are almost certainly saturated, and thus the equivalent width is not
a tracer of the gas column density. In addition, we also lack the knowledge of
the \ion{H}{1}\ column density, as well as of the ionization correction. Since
no absorption by Fe is detected, however, and since in saturated lines the
column density is underestimated, the ratio of equivalent width of
\ion{Fe}{2}\ to \ion{Mg}{2}\ provides a crude limit to the elemental abundance
ratio, if both species are contained in the same gas and partake of the same
kinematics:
$$\Big[{\hbox{Mg}\over\hbox{Fe}}\Bigr] > Log\Bigl({W_{\hbox{\ion{Mg}{2}}}\over
  W_{\hbox{\ion{Fe}{2}}}}\Bigr) \approx Log\Bigl({6.3\over 1.5}\Bigr)=0.623.$$
We can then compare this lower limit to measures of the abundance ratio of
$\alpha$--elements to iron--peak in galaxies, and in particular to bright
ellipticals, since these systems are more enriched in magnesium relative to
iron than other galaxy types (see, e.g. Thomas et~al. 2010; Jorgensen 1997;
Worthey, Faber \& Gonzales, 1992). Figure \ref{fig:Mg2Fe_abund} shows the
distribution of the ratio $[\alpha/Fe]$ for early--type galaxies with velocity
dispersion $27<\sigma<236$ \kms (red histogram, Sansom \& Northeast 2008) and
$50<\sigma<360$ (black histogram, Thomas et~al. 2005) and the lower limit to
the abundance ratio of Mg and Fe in our gas estimated above. As the figure
shows, the chemical enrichment pattern of our gas appears to significantly
differ from that observed in the galaxies. Stars in the halo of the Milky Way
also have $\alpha$--elements enhanced relative to iron (see Wheeler, Sneden \&
Truran 1989). Bonifacio et~al. (2009) find that the most metal--poor stars in
the halo of the Milky Way have $[\alpha/Fe]\lesssim 0.8$ (see their Figure 4),
which provides an estimate of the elemental ratio in nucleo--synthesis
products by Type II SNe. Clearly, being able to measure the $[\alpha/Fe]$
ratio for our gas in future spectra with higher S/N will be invaluable to
constrain its nature and the nucleo-synthetic history.

We have further investigated the Mg and Fe elemental ratio of the gas using
version 8.00 of the photo-ionization code CLOUDY (Ferland et~el. 1998) to
calculate the column density ratio of \ion{Mg}{2}\ to \ion{Fe}{2}\ if
photo--ionized by the metagalactic far--UV background. This is illustrated in
Figure \ref{fig:Mg2Fe_abund}, where the column density ratio is plotted as a
function of the ionization parameter given the QSO spectrum of Haardt \& Madau
(2005), assuming one--tenth solar metallicity for the gas chemical abundance
and using the solar elemental ratio of $[\alpha/Fe]$ by Asplund et~al. (2009)
for three assumed values of the \ion{H}{1}\ column density. We have also
repearted the calculation using the QSO+galaxy version of the Haardt \& Madau
spectrum, finding very similar results. The top abscissa in the figure shows
the value of the \ion{H}{1}\ volume density, in cm$^{-3}$, corresponding to
the ionization parameter for the case of column density $N_H=2\times 10^{20}$
cm$^{-2}$. As the curves show, if the \ion{H}{1}\ column density is
sufficiently high, e.g. like that of DLAs, most of the gas cloud is shielded
from the ionizing effects of the external radiation, and the column density
ratio of \ion{Mg}{2}\ and \ion{Fe}{2}\ remains approximately constant over a
relatively large range of the ionization parameter (this ratio can actually be
lower than the solar value in this case because some of the Mg is in the form
of \ion{Mg}{1}). For significantly lower column density, e.g. similar to and
lower than that of Lyman Limit Systems (LLS), $2\times 10^{20}$ cm$^{-2}$,
then the outer regions of the cloud become increasingly more ionized as the
ionization parameter increases and, due to the differences in their ionization
potential, the \ion{Mg}{2}/\ion{Fe}{2}\ ratio increases with it. The
conclusion of this calculation is that if the column density of the gas is
indeed as high as we have argued above, then the observed strength of the
\ion{Mg}{2}\ and \ion{Fe}{2}\ absorption features is expected to be similar,
if their elemental ratio is similar to solar, and similar to what observed in
the ISM of star--forming galaxies in the $z\sim 1.61$ overdensity. Since we do
observe \ion{Mg}{2}\ but not \ion{Fe}{2}, even if we could have observed it,
either in the combined spectrum of the LBG absorption systems or in the
combined spectrum of the GMASS background galaxies, the calculation above
strongly supports our interpretation that the absorbing gas is chemically
younger than the ISM in galaxies. Finally, as a consistency check we have
verified that if the spatial size of the absorbing cloud is of the order of
$\sim 10$ kpc, i.e. the size subtended by the angular diameter of galaxies in
range of redshift discussed here (e.g. see Ferguson et~al. 2004), then a
DLA--like column density implies \ion{H}{1}\ volume density of the order of
$2\times 10^{20}$ cm$^{-2}/3.086\times 10^{21}$ cm $=0.065$ cm$^{-3}$, namely
in the same range of density expected given the Haardt \& Madau spectrum.

\begin{figure}
\epsscale{1.0}
\plotone{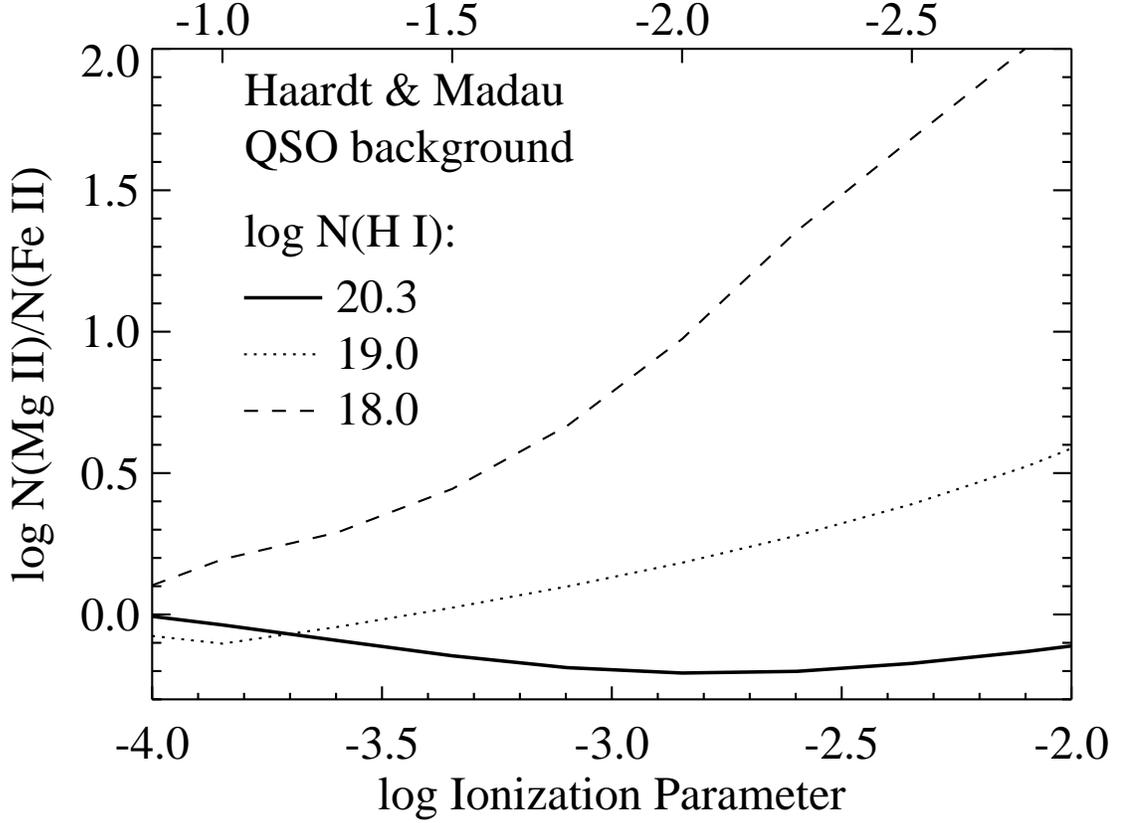}
\caption{The column density ratio of \ion{Mg}{2}\ to \ion{Fe}{2}\ plotted as a
  function of the ionization parameter for the Haardt \& Madau (2005)
  metagalactic ionizing spectrum for three assumed values of the
  \ion{H}{1}\ column density, assuming one-tenth solar abundance and using 
  solar elemental ratio of $\alpha$--element to iron--peak from Asplund
  et~al. (2009). The top abscissa shows the value of the \ion{H}{1}\ volume
  density, in cm$^{-3}$, corresponding to the ionization parameter for the
  case of column density $N_H=2\times 10^{20}$ cm$^{-2}$.
\label{fig:cloudyfig}}
\end{figure}

\subsection{Evidence Of \ion{Mg}{2}\ Gas Accreting On Some Of The Galaxies?}

In Section 3 we observed that the \ion{Mg}{2}\ absorption line found in the
co--added spectrum of the sixteen GOODS galaxies belonging to the $z\approx
1.61$ overdensity (spectrum c) of Figure \ref{fig:stack_lbgabs}) seems to
include a broad, asymmetric ``red wing'' feature extending to the red of the
\ion{Mg}{2}~$\lambda2803$ line, with possible additional discrete lines
over--imposed to it (see spectrum c) in Figure \ref{fig:stack_lbgabs}).  At
the same time, no similar blue wing is observed in the spectrum. Such an
absorption profile can result from gas along the line of sight to most of the
galaxies, located in relatively close spatial proximity to them, that is
moving toward the galaxies or toward the overdenisty as a whole, i.e. receding
from the observer. This gas could also be accreting onto some of the
galaxies. In view of the importance that such evidence would play in
constraining the nature of the gas that we have identified in the context of
cold accretion, we have further investigated this issue taking advantage of
the GMASS spectroscopic sample, which is larger than our GOODS one and
includes deeper spectra.

\begin{figure}
\epsscale{0.9}
\plotone{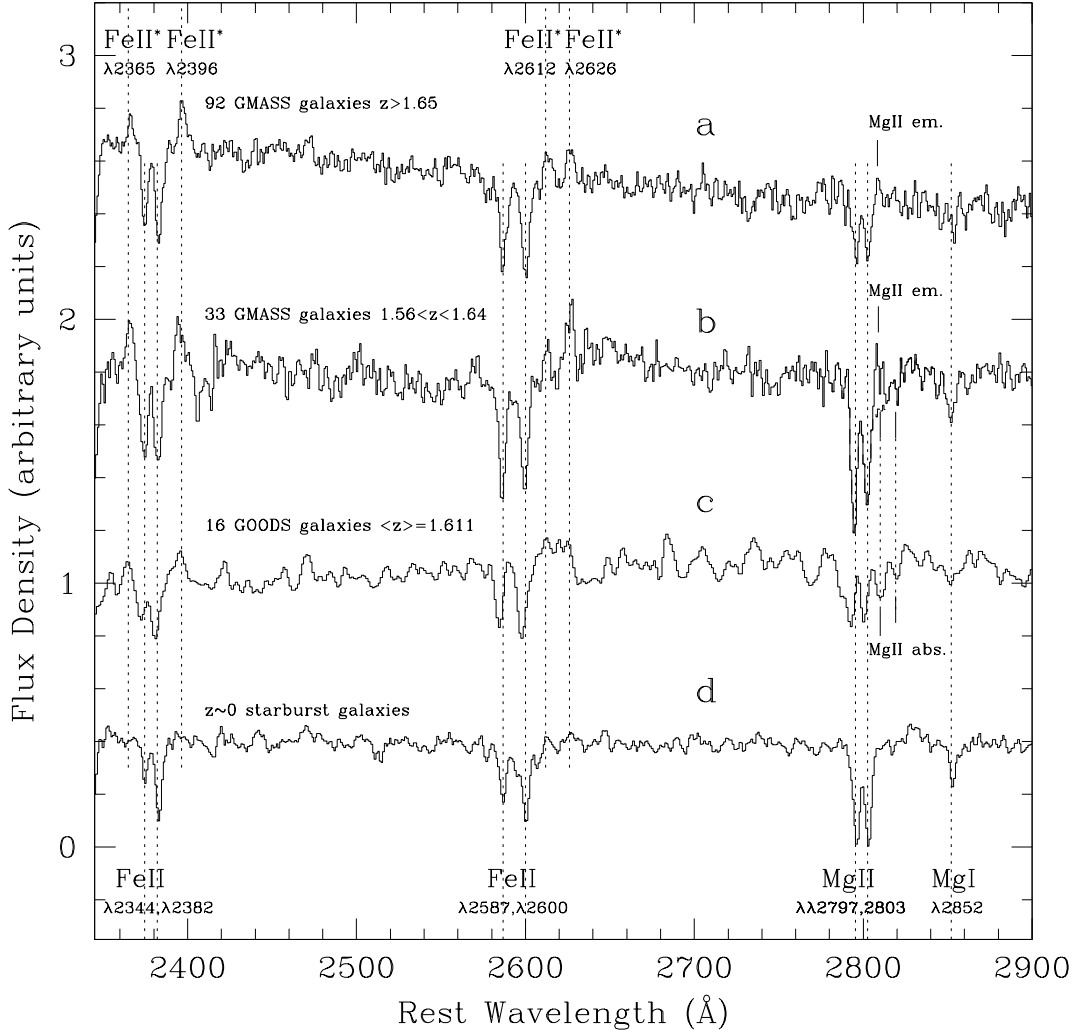}
\caption{{\bf (a)} The co--added spectrum of the 33 GMASS galaxies with
  $1.56\le z\le 1.64$ in the $z\approx 1.61$ overdensity and {\bf (b)} that of
  the 92 GMASS galaxies at $z>1.65$ in the background. Also plotted are {\bf
    (c)} the co--added spectrum of the 16 GOODS overdensity galaxies and {\bf
    (d)} that of the $z\sim 0$ starbursts, the same as those shown in Figure 
  \ref{fig:stack_lbgabs}. The GOODS stack shows the possible presence of a
  broad ``red wing'' absorption feature, extending to the red of the
  interstellar \ion{Mg}{2}\ absorption, with additional discrete components
  (marked by vertical segments) over--imposed. The GMASS stack shows a
  discrete absorption feature at approximately the same wavelength of one of
  similar discrete features in the GOODS stack. The broad ``red wing'',
  however, is not as clearly visible, mostly because of an emission line
  (marked by a vertical segment), immediately to the red of the interstellar
  \ion{Mg}{2}\ absorption feature. This is the emission component of the
  \ion{Mg}{2}\ P--Cygni feature frequently observed in star--forming galaxies
  at high redshift together with fine--structure emission lines of
  \ion{Fe}{2}$^*$ (Rubin et~al. 2010a,b; Weiner et~al. 2009), which are indeed
  detected in all of our co--added spectra.
\label{fig:accr_stack}}
\end{figure}

\begin{figure}
\epsscale{0.9}
\plotone{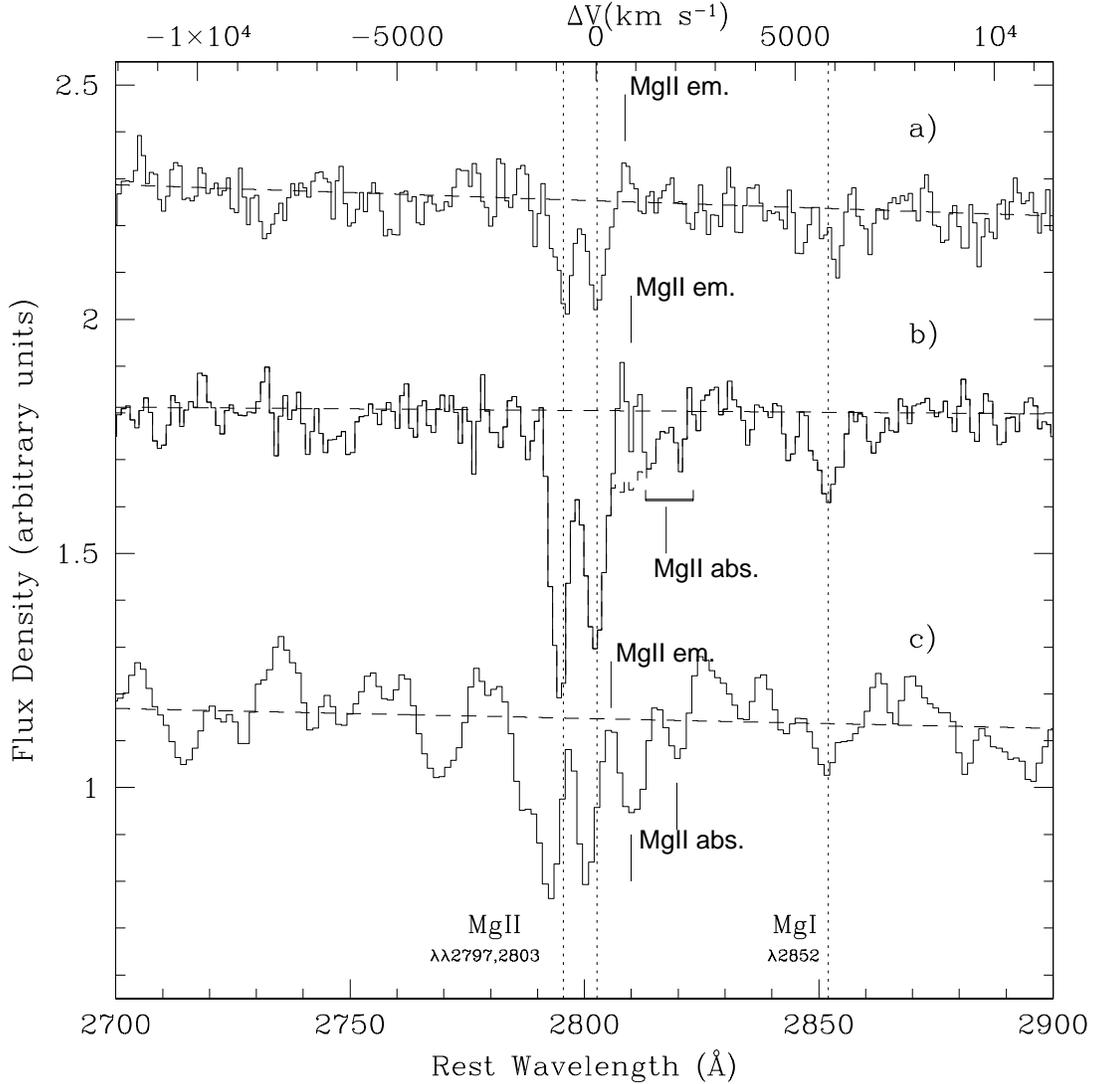}
\caption{A zoom of the spectral region around the \ion{Mg}{2}\ absorption of
  a) the GMASS co--added spectrum of the background galaxies ($z>1.65$); b)
  the GMASS and c) GOODS co--added spectra of the galaxies in the $z\sim 1.61$
  overdensity. Vertical dashed lines mark the wavelength of the
  \ion{Mg}{2}$\lambda\lambda 2797,2803$ doublet, and the top horizontal axis
  plots the velocity difference from the \ion{Mg}{2}~$\lambda2803$
  component. The presence of outflows is observed in in spectra b) and c) as a
  blueshift of the \ion{Mg}{2} doublet. A linear spline fit to the continuum
  in two wavelength intervals bracketing the \ion{Mg}{2}\ region of each
  spectrum is also shown. Short vertical segments above the spectra mark the
  position of the residual emission components of \ion{Mg}{2}\ P--Cygni
  features diluted by the interstellar absorption and the relatively low
  spectral resolution. The dashed curve at the bottom of spectrum b) shows the
  continuum after a gaussian fit to the \ion{Mg}{2}\ emission line has been
  subtracted, illustrating the presence of a broad ``red wing'' absorption
  feature. Such a ``red wing'' broad absorption also seems present in spectrum
  c). Short vertical segments below spectra b) and c) mark likely
  \ion{Mg}{2}\ discrete absorption features within the ``red wing''
  feature. No such ``red wing'' absorption features is observed in spectrum
  a).
\label{fig:accr_stack_zoom}}
\end{figure}

We have averaged together the GMASS rest--frame spectra of 33 galaxies
belonging to the $z\approx 1.61$ overdensity ($1.56\le z\le 1.64$, $\langle
z\rangle=1.608$) to provide an independent average spectrum of galaxies in the
same structure where we have detected our ``pristine'' \ion{Mg}{2}\ gas, whose
equivalent exposure time of $T_{exp}=855$ hr provides significantly higher
sensitivity to gas absorption features than any individual galaxy. As a
control spectrum, we have averaged together the 92 GMASS spectra of galaxies
at $z>1.65$, i.e. in the background of the overdensity this time after
shifting each one of them to the rest frame. It makes no difference if the
spectra are scaled prior to computing the average. Whenever possible, the
redshift of the galaxies is measured from the [\ion{O}{2}] nebular emission
lines, since this provides a good measure of the systemic redshift. When
[\ion{O}{2}] is not present in the available spectral range, the redshift is
estimated from the \ion{Mg}{2}\ and \ion{Fe}{2}\ absorption lines; 19 of the
33 overdensity galaxies and 1 of the background galaxies have [\ion{O}{2}]
redshift. We have also experimented with co--adding only galaxies with
[\ion{O}{2}] emission and those without, finding similar results, as we will
discuss later. The 33 co--added spectra are shown in Figure
\ref{fig:accr_stack} together with the GOODS stack of the overdensity galaxies
and the stack of the $z\sim 0$ starburst galaxies, for comparison (the latter
two spectra are the same as those shown in \ref{fig:stack_lbgabs}). Figure
\ref{fig:accr_stack_zoom} zooms in the spectral region around the
\ion{Mg}{2}\ absorption in the GMASS and GOODS stacks and also shows the
velocity difference of the absorption features relative to the rest--frame
wavelength of the \ion{Mg}{2}~$\lambda 2803$ line. Again, possible discrete
absorption features to the red of the main interstellar absorption features
are marked by vertical segments.

Like the GOODS, the GMASS stack of the $z\approx 1.61$ overdensity galaxies
also shows a discrete absorption feature to the red of the
\ion{Mg}{2}~$\lambda2803$ line, apparently characterized by two minima, at
approximately the same wavelength as the redder feature of the two observed in
the GOODS stack. In fact, the redder of the two minima, marked by the right
vertical segment in Figure \ref{fig:accr_stack_zoom}, coincides (within the
resolution of the spectra) with the redder feature of the GOODS spectrum.
Overall, the whole absorption feature, i.e. including the two contiguous
minima, is detected at the $\gtrsim 10\sigma$ level, similar to the
\ion{Mg}{1}~$\lambda2852$ line. The GMASS stack, however, shows a relatively
strong emission line (more on this later) immediately to the red of the strong
interstellar \ion{Mg}{2}~$\lambda2803$ absorption line, roughly at the same
wavelength of the bluer of the two discrete absorption features in the GOODS
stack, as illustrated in Figures \ref{fig:accr_stack} and
\ref{fig:accr_stack_zoom}. Overall, although taken individually each
absorption feature is detected at low level, their simultaneous presence in
both the GMASS and GOODS co--added spectra and the approximated coincidence in
wavelength strongly argues in favor of them being real, since the GMASS and
GOODS galaxies come from two different samples and cover somewhat different
regions of the GOODS--S field (see Figures \ref{fig:overdens_dahlen} and
\ref{fig:overdens_guo}). No similar absorption features are observed in the
co--added spectrum of the 92 background galaxies at $z>1.65$. Also, neither
the two GMASS stacks nor the GOODS one show any obvious similar absorption
feature to the blue of the main \ion{Mg}{2}~$\lambda2797$ interstellar line.

All these absorption features, both the discrete ones and the broader,
continuum one, must be caused by a distribution of intervening
\ion{Mg}{2}\ clouds. An obvious interpretation for their redshift relative to
the \ion{Mg}{2} doublet and for the broad profile of the broad ``red wing''
absorption profile is that the gas is moving toward the overdenisty and/or at
least some of the galaxies contributing to the two stacked spectra. An
extended \ion{Mg}{2}\ cloud would produce the broad continuous ``red wing''
absorption profile observed in their co--added spectrum, while individual,
denser clouds embedded in this medium would be responsible for the discrete
features. The top horizontal scale in Figure \ref{fig:accr_stack_zoom} shows
the velocity difference relative to the $\lambda2803$ line of the
\ion{Mg}{2}\ doublet.  Thus, the implied relative velocity of the two discrete
features in the GOODS stack (spectrum c) in Figure \ref{fig:accr_stack}) would
be $\approx 700$ and $\approx 1600$ \kms, respectively. Such values of the
velocity are higher than the predictions for accretion of cold gas onto $M\sim
10^{12}$ halos, which are about $200$ \kms, but they are at least
qualitatively consistent with the average relative velocity between gas clouds
and galaxies within the much more massive overdensity. We also recall, in this
regard, that the redshift of the four intervening LBG absorption systems is
displaced by up to $\pm 1000$ \kms\ relative to the peak of the overdensity.

The broad ``red wing'' is analogous to the ``blue wing'' absorption profile
observed immediately to the blue of the far--UV interstellar metal absorption
lines discussed by Steidel et~al. (2010) in the same types of galaxies. In the
case of the ``blue wing'', the feature originates in gas outflows from the
galaxies, while the ``red wing'' discussed here originates in inflows. Since
ultra--strong \ion{Mg}{2}\ absorption traces hydrogen gas with $N_H\gtrsim
10^{20}$ cm$^{-2}$, we interpret the detection of the ``red wing'' as evidence
of accretion of cold, chemically relatively young gas onto the $z\approx 1.61$
overdensity. This gas might very well feed star formation in the star--forming
galaxies that belong to the overdensity, although there is no direct evidence
of this fact.

The broad ``red wing'' is not as clearly observed in the GMASS stack of the
overdensity galaxies as it is in the GOODS one because of the presence of a
relatively strong emission line, which is the emission component of a
\ion{Mg}{2}\ P--Cygni feature. As noted before, Figure \ref{fig:accr_stack}
and Figure \ref{fig:accr_stack_zoom} show the presence of emission lines
immediately to the red of the $\lambda=2803$ \AA\ component of the
\ion{Mg}{2}\ doublet. Emission of \ion{Mg}{2}\ is very rarely seen in
starburst galaxies in the local universe that have no AGN activity
(e.g. Tololo 1924-416, Kinney et~al. 1993), but it appears to be more common
at high redshift (Weiner et~al. 2009; Rubin 2010a,b,c), where the feature is
observed in conjunction with emission lines by fine--structure transitions of
\ion{Fe}{2}$^*$ and has a P--Cygni profile. We detect the \ion{Fe}{2}$^*$
fine--structure lines in all our co--added spectra (in fact, they are detected
in the majority of the individual spectra), although they do not seem to be
present in the $z\sim 0$ spectrum. While observational evidence of these lines
is accumulating at high redshift (Weiner et~al. 2009; Rubin et~al. 2010a,b;
Kornei et~al. in preparation) very little is known about the physical
mechanisms that produce them. Rubin et~al. (2010c) and Prochaska, Kasen \&
Rubin (2011) have proposed that both the \ion{Fe}{2}$^*$ fine structure
emission lines and the P-Cygni \ion{Mg}{2}\ feature originate by photon
scattering in the outflows that characterize distant star--forming galaxies
(Weiner et~al. 2009; Steidel et~al. 2010), although these lines can also be
produced in recombination regions (Kinney et~al. 1993) or AGN (Vestergaards \&
Wilkes 2001). The \ion{Fe}{2}$^*$ lines are very well detected in two of the
GMASS stacks and in the GOODS stack shown in Figure \ref{fig:accr_stack} (see
also the stack of the GOODS spectra of the overdensity galaxies in Figure
\ref{fig:stack_lbgabs}), where the two lines \ion{Fe}{2}$^*~\lambda2612$ and
\ion{Fe}{2}$^*~\lambda2626$ are blended because of the lower spectral
resolution. Interestingly, all the \ion{Fe}{2}$^*$ lines and the
\ion{Mg}{2}\ emission line are stronger in the GMASS stack of the 33
overdensity galaxies than in the stack of 92 background ones, despite the
longer equivalent exposure time of the latter, very likely because for nearly
all of them [\ion{O}{2}] is outside of the available spectral range, and thus
their redshift is determined from the absorption features of \ion{Fe}{2}\ and
\ion{Mg}{2}. Since these lines are heavily affected by the outflows, their
redshift deviates from the systemic one by an random amount, between a few to
several hundred km/s (Steidel et~al. 2010), and thus the strength of the stack
of the \ion{Fe}{2}$^*$ lines, which are at the systemic (Rubin et~al. 2010c),
is diluted. In comparison, more than half of the redshifts of the $z\sim 1.61$
overdensity galaxies are measured from the [\ion{O}{2}] line.

As noted before, the \ion{Fe}{2}$^*$ lines are not observed in the stacked
spectrum of the $z\sim 0$ starburst galaxies by Leitherer et~al. (2010)
reproduced in Figures \ref{fig:stack_lbgabs} and \ref{fig:accr_stack}. We
suggest that this is due to the fact that UV spectroscopic observations of
nearby galaxies have mostly focused on bright, relatively isolated regions of
star formation, and not on the integrated light of the galaxies. Combined with
the relatively small apertures of the space--born spectrometers and the large
apparent diameter of the sources, this fact has likely resulted in most of the
spectra not sampling the volumes of space affected by the outflows.  Some of
the \ion{Fe}{2}$^*$ lines also seem weak in the 7--galaxies stacked spectrum
at $\langle z\rangle\sim 1.9$ shown in Figure \ref{fig:stack_lbgabs}, but this
is due to the relatively low S/N of the individual spectra and to the small
number of galaxies entering this stack.

The fact that the full P--Cygni profile, i.e. both absorption and emission
components, is not clearly recognized in our spectra (but it is very well
detected in the spectra by Rubin et~al. 2010a,c), especially the emission
component of the $\lambda 2796$ \AA\ transition of the \ion{Mg}{2}\ doublet,
should not come as a surprise in our relatively low resolution
spectra.\footnote{The resolution of the Keck/LRIS spectra Rubin et
  al~al. (2010a,c), where the P--Cygni profile is clearly detected in the
  spectra of individual galaxies, is $\approx 2.5\times $ higher than that of
  our spectra. That of the Keck/DEIMOS spectra by Weiner et~al. (2009), who
  also observe the P--Cygni profile for some of the galaxies, is $\approx
  10\times $ larger.} As reported by Rubin et~al. (2010a,b,c) both the
fine--structure Fe$^*$ lines and the P--Cygni \ion{Mg}{2}\ feature {\it are
  observed approximately at the systemic redshift}. Thus, the combined effect
of the relatively low spectral resolution, the strength and large velocity
width of the interstellar absorption, and the fact that a large number of
galaxies with different outflow velocities enter the stack (recall that the
spectra are put on the rest--frame using the redshift of the
[\ion{O}{2}]~$\lambda 3727$ nebular line, which traces the systemic redshift
of the galaxies, only when the feature is available in the spectral range) is
that the emission component of the P--Cygni profile is diluted by the strong
interstellar absorption.  Because the generally large velocity of the
outflowing gas blue--shifts the absorption component, some of the
\ion{Mg}{2}~$\lambda2803$ P--Cygni emission that is located at the systemic
redshift survives, and this is what we are most likely observing. The emission
line in the GMASS spectrum seems to exhibit an absorption feature, which, as
we noted previously, is observed at the same wavelength as the red absorption
feature in the red wing of the GOODS spectrum. This apparent trough could give
the GMASS \ion{Mg}{2}\ emission the resemblance of the double--peaked
line. The emission component of the \ion{Mg}{2}\ P--Cygni feature is also
possibly detected in the GOODS co--added spectrum (see Figures
\ref{fig:accr_stack} and \ref{fig:accr_stack_zoom}). Finally, we note that a
virtually identical spectral morphology of the \ion{Mg}{2}\ P--Cygni profile
is observed in the in the {\it HST}/FOS UV spectra of Tololo 1924-416 by
Kinney et~al. (1993), which have similar resolution ($R\approx 400$).

Subtracting the \ion{Mg}{2}\ emission from the spectrum makes the broad ``red
wing'' absorption profile evident in the GMASS co--added spectrum too, showing
that it extends to approximately the same velocity separation from
\ion{Mg}{2}~$\lambda2803$ as the one in the GOODS spectrum, namely $\Delta
V\lesssim 2000$ \kms. This is illustrated in Figure \ref{fig:accr_stack_zoom},
where the dashed line shows the residual continuum after a gaussian profile
has been fitted to the emission line and subtracted. The Figure shows a zoom
in of the spectral region around the \ion{Mg}{2}\ doublet in the co--added
spectra of the GMASS background galaxies ($z>1.65$) and of the 33 GMASS and 16
GOODS $z\sim 1.61$ overdensity galaxies. A linear spline fit to the continuum
in two spectral regions that bracket the \ion{Mg}{2}\ doublet
($2650\le\lambda\le2750$ \AA\ and $2830\le\lambda\le2930$ \AA) is also shown
for each spectrum to help identify the absorption features. The broad ``red
wing'' absorption feature in the GMASS spectrum of the overdensity galaxies
results in part from the removal of the emission line, but also, in large
part, from the presence of what look like absorption features immediately to
the red of the emission line, marked by a horizontal square bracket in the
Figure \ref{fig:accr_stack_zoom}. The discrete absorption lines possibly
detected in the GOODS co--added spectrum also are marked in the figure.

\begin{figure}
\epsscale{0.9}
\plotone{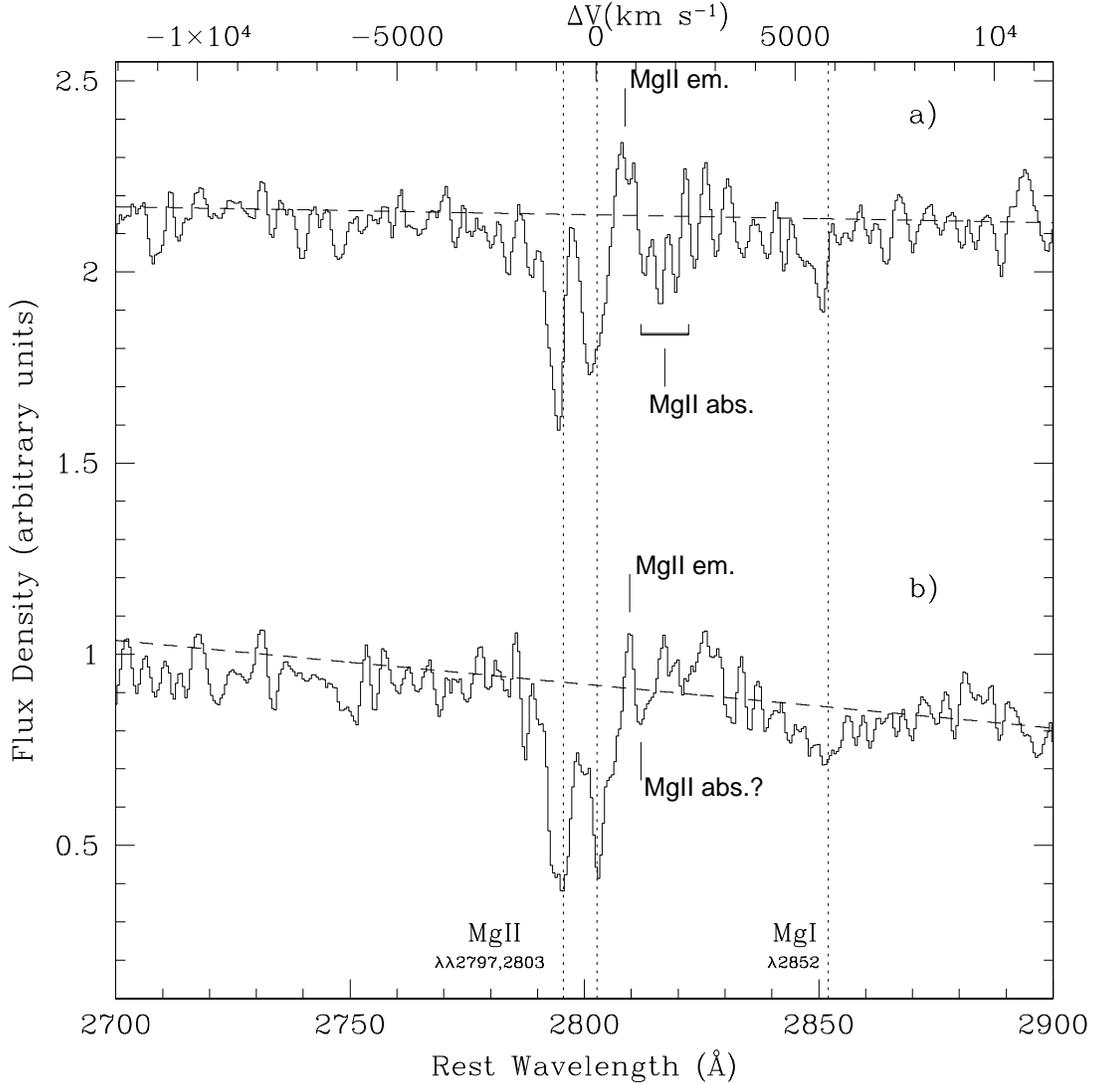}
\caption{Co--added spectra of GMASS galaxies in the $z\sim 1.61$
  overdensity. The top spectrum (a) includes galaxies with redshift measured
  form the [\ion{O}{2}] nebular emission line, while the bottom spectrum (b)
  galaxies whose redshift comes from interstellar absorption lines such as
  \ion{Fe}{2}\ and \ion{Mg}{2}. Spectrum (a) has a stronger emission
  component of the \ion{Mg}{2}\ P--Cygni feature than spectrum b) of Figure
  \ref{fig:accr_stack_zoom}, but it is otherwise quite similar to it,
  including the presence of absorption right to the red of the emission
  line. Removal of the emission line would also result in a ``red wing''
  absorption feature. A possible broad ``blue wing'' absorption profile is
  possible observed immediately to the blue of the \ion{Mg}{2}~$\lambda2797$
  absorption line, too. Broad ``red wing'' absorption is less clear in
  spectrum (b); while the removal of the \ion{Mg}{2}\ emission line would make
  it more evident, the absorption profile would not be as extended as in
  spectrum(a).
\label{fig:accr_stack_zoom_test}}
\end{figure}

We emphasize that no ``red wing'' absorption feature seems to be present in
the co--added spectrum of the GMASS background galaxies, namely spectrum a) in
Figure \ref{fig:accr_stack}, even if the (weak) emission component of the
\ion{Mg}{2}\ P--Cygni feature is subtracted. About three times more galaxies
are included in this spectrum than in that of the overdensity galaxies. Thus,
it seems unlikely that in the stack of the overdensity galaxies the feature is
the result of some systematics arising during the averaging procedure. For
example, one could imagine that deviations of the measured redshift of the
galaxies from their systemic value might produce the feature by broadening the
galaxies' interstellar \ion{Mg}{2}~$\lambda2803$ absorption line. Random
deviations up to several hundred \kms\ will occur for those galaxies where the
[\ion{O}{2}] nebular emission line is not available and the redshift is
measured from the Mg and Fe interstellar absorption lines, which are typically
blushifted relative to the systemic redshift due to the presence of outflows
(Steidel et~al. 2010). The co-added spectrum of the background galaxies, whose
redshift are essentially all measured from the interstellar lines suggests
that this is not the case. We have further investigated this possibility by
separately co--adding the spectra of the GMASS overdensity galaxies with
redshift measured from the [\ion{O}{2}] emission line (19 spectra) and of
those with redshift measured from the absorption lines (14 spectra), which are
reproduced in Figure \ref{fig:accr_stack_zoom_test} as spectrum a) and b),
respectively.  Spectrum a) has a stronger \ion{Mg}{2}\ P--Cygni emission
component, but is otherwise qualitatively very similar to the full stack,
i.e. spectrum b) of Figure \ref{fig:accr_stack_zoom}, with a clear depression
of the continuum right to the red of the emission line. Removing the emission
line would result in the same ``red wing'' absorption profile like the one we
have illustrated in \ref{fig:accr_stack_zoom} for the full--stack spectrum. We
also note that a ``blue wing'' absorption is now possibly observed in spectrum
a); this would be the effect of absorption in the outflows, quite similar to
what observed in other UV low--ionization metal absorption lines (see Steidel
et~al. 2010).  The ``red wing'' absorption in spectrum b) seems to be much
weaker. The emission component of the \ion{Mg}{2}\ P--Cygni feature is also
weaker than in spectrum a), and while removing the feature would result in a
broad ``red wing'' absorption, this would not reach the same extent as in the
case of spectrum a). The ``blue wing'' due to the outflows, if present, also
is very weak. Finally, because the redshift of the individual spectra entering
in spectrum a) is very close to the systemic one, it can be seen that both the
\ion{Mg}{2}\ and \ion{Mg}{1}\ absorption lines are blueshifted, by about 250
\kms, due to the presence of outflows. Obviously, this effect cannot be
observed in spectrum b) because the individual spectra are placed in the rest
frame using the redshift of the absorption lines themselves.

It is important to realize that, although subtracting the emission component
of the \ion{Mg}{2}\ P--Cygni feature is a necessary step if one wants to
explore gas accretion in quantitative details, we have little guidance on how
to objectively carry out the subtraction procedure at this time; our
subtraction of a gaussian fit to the line remains arbitrary. As we have
mentioned earlier, the \ion{Mg}{2}\ P--Cygni feature in star--forming galaxies
at high redshift is observed in conjunction with the fine structure emission
lines of \ion{Fe}{2}$^*$, but not much else is known about these features and
their empirical phenomenology and systematics, e.g. how the line strength of
the two species depend on each other and on the properties of the galaxies
and/or of the outflows. We are only aware of the photon scattering mechanism
proposed by Rubin et~al. (2010c) and Prochaska, Kasen \& Rubin (2011) to
explain these lines at this time. We have also checked for the possibility
that some of the galaxies in our sample might host an AGN, but none is
individually detected in the recently released ultra--deep Chandra X--ray
image ($T_{exp}=4$ Msec, see Gilli et~al. 2011) in the CDFS. With no guidance
from either the theory or the observations, it is not possible to go beyond
the qualitative and crude analysis discussed here. Rubin et~al. (2010b) follow
a similar procedure of fitting and subtracting the emission component of the
\ion{Mg}{2}\ P--Cygni feature (to derive the equivalent width of the
\ion{Mg}{2}\ interstellar absorption) using a five--parameter model that
separately accounts for the emission and absorption components. As they
discuss, this model does not produce acceptable results for co--added spectra,
since it tends to be driven by noise features in the line profiles. Thus, we
have decided to avoid the further complication of the model and simply remove
the feature after fitting the ``local'' line profile with a gaussian, since
this seems to give comparable results, and our initial goal is to explore the
{\it the existence} and qualitative aspects of cold gas accretion.

Regardless of the details, however, even a very conservative subtraction of
the line from the co--added spectra shown in figures \ref{fig:accr_stack_zoom}
and \ref{fig:accr_stack_zoom_test} (such as simply eliminating the local
excess flux) results in an obvious ``red wing'' absorption profile that
extends from immediately to the red of the \ion{Mg}{2}\ interstellar feature
to $\Delta V\approx 2000$ \kms\ from the systemic wavelength of the 
\ion{Mg}{2}~$\lambda2803$ line.

The ``red wing'' absorption profile is observed in both the GMASS and GOODS
co--added spectra of galaxies in the $z\approx 1.61$ overdensity but is not
observed in the much deeper stacked spectrum of the 92 background galaxies at
$z>1.65$ in the background, implying that its detection is related to the
presence of the overdensity itself. This suggests that the average covering
factor of gas around galaxies that belong to the overdensity is substantially
larger than that around isolated field galaxies. We plan to return on this
issue more quantitatively in a upcoming paper.

\section{Discussion And Conclusions}

Three key results have emerged from this work. The first is the identification
of substantial amount of ``cold'' gas, i.e. $T\approx 10^4$ K, in an
overdensity of galaxies at $z\approx 1.61$ that is not directly associated
with (i.e. resides in the gaseous halo of) bright galaxies, but rather is
either ``circum'' or ``intra'' overdensity gas, or both. The gas appears to be
chemically younger than the outflows from star--forming galaxies at the same
redshift (including the galaxies in the overdensity), which have similar
physical conditions (e.g. Tremonti et~al. 2007). In other words, the gas is
part of the structure and is spatially distributed around and/or within it but
it is not part of the galaxies that belong to the structure. This gas has been
identified by the \ion{Mg}{2}\ absorption that it imprints in the spectra of
background LBG at $z\sim 3$, by the absorption feature in the co--added
spectra of a large number of galaxies in the background of the overdensity, as
well as in the co--added spectrum of galaxies within the overdensity.

We can derive a crude estimate of the average covering factor of the
high--column density gas from the ratio of detected intervening ultra--strong
\ion{Mg}{2}\ absorption systems, i.e. the 4 systems in the spectra of G1, G2
and G3, to the number of spectra that have sufficient S/N to detect such
systems, namely the 21 spectra of GOODS LBGs and the 92 spectra of GMASS
background galaxies. We find $C_F\sim 4/105\approx 4$\%. This is consistent
with recent current theoretical estimates for isolated galaxies hosted in
$M\sim 10^{12}$ \msun\ halos, i.e. $C_F\lesssim 3$\% at $z\sim 1.5$, and given
the fact that our spectra are only sensitive to absorption with large
equivalent width, probably even consistent with the expected trend of
increasing cold gas covering factor with larger mass, i.e. $C_F\approx 10$\%
around halos with $M\sim 10^{14}$ \msun\ at the same redshift (Kimm
et~al. 2010). The absorption features observed in the co--added spectra of the
background galaxies and of the galaxies in the overdensity suggest the
presence of both continuous absorption as well as absorption by discrete
clouds, consistent with a projected gas distribution where structures capable
of higher column density are embedded in a more diffuse
medium. Observationally, the probability that gas clouds intersect the line of
sight to galaxies should be larger in an overdensity than for background
galaxies, since the surface density of the former is enhanced, by about
six-fold in our case, than the latter. Kimm et~al. (2010) suggest that the low
covering factor around isolated galaxies, coupled with the fact that the
interstellar absorption lines are strong and broad in velocity space, are the
key factors behind the current lack of detections of cold accretion. Thus,
while waiting for the sensitivity of 30--meter telescopes, targeting galaxy
overdensities seems to be a more efficient way to make some progress.

Essentially no information is available about the spatial morphology of the
high--column density gas, i.e. whether filamentary or not, as predicted by the
simulations (e.g. Keres et~al. 2005; Dekel et~al. 2009). Curiously, the three
LBG with the intervening \ion{Mg}{2}\ absorption systems are aligned along a
straight line and the spatial separation and velocity difference between the
four LBG \ion{Mg}{2}\ absorbers are consistent with an elongated distribution
of the gas: the velocity difference between G1's absorber and the first one of
G2, $\Delta V\approx 500$ \kms, and their projected spatial separation at
$z\approx 1.61$, $D\approx 200$ kpc, are consistent with them being produced
by the same gas mass. A homogeneous cloud would have to be either more massive
than $\approx 2.5\times 10^{10}$ \msun, the same mass of an $\approx L^*$
galaxy at the same redshift (Reddy et~al. 2009), if roughly spherically
symmetric, or be preferentially distributed along the separation between the
two absorbers, unless $\langle C_F\rangle\ll1$ and we have accidentally hit
two spots of high column density. Similarly, the projected separation between
the second absorber of G2 and that of G3 is $\approx 1$ Mpc, and their
velocity difference is $\approx 800$ \kms. If this velocity difference were
due to pure Hubble flow, then the physical distance between the two troughs
would be $d=\Delta V/H(z)=4.6$ Mpc, where the Hubble constant $H_z$ at
$z=1.61$ is $H_{1.61}=172$ \hbl. But these two alignments can very well be
just a fortuitous coincidence. If the geometry of the high--column density gas
mass is filamentary, the finite size of the LBG yields a lower limit to the
linear mass density along the filament, which is $m_F>0.8\times 10^8 \times
{\cal C}_F\times (N_H/ 10^{20}{\rm cm}^{-2})$ \msun~(10~kpc)$^{-1}$, where
${\cal C}_F$ is the gas average linear covering factor along the filament
(which is different form the average volume covering factor $C_F$ discussed
above). If this mass accretes onto some galaxy, as the spectra of galaxies in
the $z\approx 1.61$ overdensity suggest, and is converted into stars, then a
lower limit to the star--formation rate that the filament can sustain is
$$SFR\approx 25\times {\cal C}_F\times\Bigl({N_H\over
  10^{20}~\hbox{cm$^{-2}$}}\Bigr) \times\Bigl({V_T\over
  300\hbox{~\kms}}\Bigr)\hbox{~~~\sfr},$$ 
where $V_T$ is the tangential velocity of the gas flow along the filament.
This lower limit is in agreement with the star--formation rate of galaxies at
$1<z<2$ (Daddi et~al. 2007; Reddy et~al. 2009). The main uncertainties here
are the width of the putative filaments and the value of ${\cal C}_F$. It is
very difficult to make more informative statements without knowing more about
the large--scale geometry of the gas. The analysis of a larger number of
background probes will bring improvement in this area.

All four intervening LBG \ion{Mg}{2}\ absorption systems have redshift located
in the wings of the redshift distribution of the overdensity, as if they
tended to ``avoid'' the galaxies. It is possible that this is, at least in
part, an observational bias due to the presence of night sky emission lines
(albeit week ones) at the wavelengths of \ion{Mg}{2}\ at the redshift of the
peak. While one could certainly imagine the possibility of astrophysical
processes that suppress high--column density gas clouds inside large
concentrations of galaxies, it is hard to reach any conclusion with only four
data points. Note, however, that large concentrations of galaxies, such as
ours, marked by spikes in the redshift distribution, do not have properties
commonly associated with galaxy clusters, although they might evolve into one
(e.g. Kurk et~al. 2009). Their spatial extent is much larger, by at least by
one order of magnitudes, than clusters and thus their galaxy density is lower
by about three orders of magnitude; no diffuse X--ray emission is observed
from them, as we directly verified in our case using the ultra--deep Chandra
X-ray images in the CDFS. In any case, however, the existence of high--column
density gas clouds around such structures seems real, with velocity
separations that can reach up to $\sim 2000$ \kms, as evidenced not only by
the LBG absorption systems, but also by the absorption features observed in
the GMASS co--added spectra of both background and overdensity galaxies. The
observation of the ``red wing'' absorption profile seems to suggest that this
might very well be gas associated with cold accretion onto the structure.

Information on the kinematics of the absorbing gas is minimal from the current
low resolution and, in the case of the four LBG absorption systems, low S/N
spectra. Since the absorption lines are very likely saturated, their velocity
width mostly reflects the velocity spread of the gas along the line of sight,
which is less than a few hundred \kms. While this is certainly consistent with
the general idea that the gas in cold streams is expected to be kinematically
``cold'' (the line broadening contributed by the temperature of the gas,
$T\sim 10^4$, is small, of the order of $\sim 10$ \kms), it offers little
insight into the dynamics of the gas, especially since we do not know much
about the spatial geometry of the absorbing troughs.

The second key result is that the cold, $T\sim 10^4$ K gas appears to be
chemically more pristine than the ISM and the gas outflows from galaxies at
similar redshift, including those galaxies that belong to the overdensity.
This evidence comes from the lack of absorption by Fe (i.e. the absorption
lines \ion{Fe}{2}~$2587$ and \ion{Fe}{2}~$2600$) in the co--added spectrum of
the absorption systems detected in the spectra of G1, G2 and G3, and in the
co--added spectrum of the 92 GMASS galaxies in the background of the
overdensity. Interstellar absorption lines by \ion{Fe}{2}\ and
\ion{Mg}{2}\ typically have similar strength at both low and high redshift, as
documented by numerous works (e.g. Leitherer et~al 2010; Weiner et~al. 2009;
Rubin et~al 2010a,b) and as illustrated here by the stacked spectra of both
local and high--redshift galaxies. All our co--added spectra of galaxies have
both Mg an Fe absorption lines with approximately equal strength, but the
co--added spectra of the four LBG absorption systems and that of the 92 GMASS
$z>1.65$ background galaxies (non registered to the rest frame) only have
\ion{Mg}{2}\ absorption. A crude lower limit to the [$\alpha$/Fe] ratio
derived from the ratio between the observed \ion{Mg}{2}\ equivalent width and
the upper limit to the \ion{Fe}{2}\ one shows that the gas has a different
enrichment pattern than that observed in galaxies. A calculation made with
CLOUDY shows that the column density ratio of \ion{Mg}{2}\ to \ion{Fe}{2}\ is
expected to be around unity, assuming a medium with one--tenth solar
metallicity, solar elemental ratio, DLA--like \ion{H}{1}\ column density and
the Haardt \& Madau (1997) spectrum of metagalactic ionizing radiation. Thus,
our finding suggests that the cold gas we have identified has a deficiency of
Fe--peak relative to $\alpha$ elements, implying that it is very likely
chemically younger and underwent a different enrichment history, than the ISM
of galaxies at the same redshift, including the outflows from star--forming
ones. A possibility is that this is relatively pristine cosmic gas,
pre--enriched by earlier generations of stars than those forming in the
galaxies of the $z\approx 1.6$ overdensity. The overabundance of
$\alpha$--elements over iron--peak would result from core--collapse (Type II)
supernovae preferentially enriching the IGM relative to Type Ia. Whether this
overabundance of $\alpha$--elements over Fe is due to the time delay between
the occurrence of Type Ia relative to Type II events or to a top--heavy IMF,
or both, cannot be addressed here.

Finally, the third key result of this paper is the possible direct detection
of accretion of cold gas on the galaxies of the overdensity or on the
overdenisty itself. This evidence, which we regard as tentative at this time,
comes from the presence of a broad ``red wing'' absorption profile in both the
GMASS and GOODS co--added spectra of galaxies that belong to the overdensity,
i.e. a broad absorption feature that extends for about $\lesssim 2000$
\kms\ red--ward of the $\lambda2803$ component of the \ion{Mg}{2}\ doublet. The
``red wing'' absorption profile is detected, albeit at low significance, in
the GOODS co--added spectrum of 16 galaxies of the overdensity. It is detected
with better significance in the GMASS co--added spectrum of 33 galaxies (the
equivalent of $\approx 855$ hours of exposure time), but only after the
removal of the emission component of a \ion{Mg}{2}\ P--Cygni feature, which
therefore needs to be understood and characterized from an empirical point of
view.

The emission line from the \ion{Mg}{2}\ P--Cygni feature, together with
emission lines from fine--structure transitions of the excited ion
\ion{Fe}{2}$^*$, seem to characterize the mid--UV spectra of star--forming
galaxies at high redshift, possibly the result of their powerful outflows (see
Rubin et~al. 2010a,b,c; Prochaska et~al. 2011), although their physics and
phenomenology remain poorly constrained at the present. They also seem to be
very common. We detect the \ion{Mg}{2}\ emission component and the
\ion{Fe}{2}$^*$ emission lines in the GOODS and GMASS co--added spectra of
galaxies that belong to the $z\sim 1.61$ overdensity and of galaxies in the
background, at $z>1.65$.

The \ion{Mg}{2}\ P--Cygni emission component clearly needs to be accounted for
and removed when looking for the signature of accretion of gas capable of
\ion{Mg}{2}\ absorption, which can manifest itself as an extended absorption
feature redshifted relative to the systemic redshift of the wavelengths of the
\ion{Mg}{2}\ doublet (the observed doublet is actually very often blueshifted
relative to the systemic redshift because of the outflows). The quantitative
details of the removal procedure, however, remain unclear, because so little
is known about the physical mechanisms responsible for this emission and thus
we do not have theoretical guidance on how to model and subtract the
feature. Regardless, however, a conservative removal of the emission feature,
i.e. simply eliminating the excess flux over the adjacent continuum emission,
already yields a ``red wing'' absorption profile in the GMASS spectrum, which
is qualitatively and quantitatively similar to that observed in the GOODS
one. Thus, while it seems prudent to regard the detection of accretion as
tentative, the simultaneous presence of the ``red wing'' in both the GMASS and
GOODS co--added spectra makes the case compelling.

In summary, we have identified substantial amounts of optically--thick, cold
($T\sim 10^4$ K) gas associated with a large overdensity of galaxies at
$z\approx 1.61$, which is not part of the extended halos of any massive
galaxy. The gas is chemically ``younger'', namely significantly more depleted
of Fe relative to Mg, than the interstellar medium and the outflows of
galaxies at the same redshift, including the ones in the overdensity. There is
evidence from the co--added spectra of star--forming galaxies that belong to
the overdensity that the gas is accreting onto them or onto the overdensity as
a whole. Crude estimates of the star formation rate that this accretion can
sustain (in fact, lower limits) based on the (scant) information we have on
the column density of the gas, its projected spatial distribution, and its
kinematics, are consistent with the measures of the galaxies' star
formation rate. We believe this is the most compelling direct observational
evidence to date in favor of the so called ``cold accretion''.  The key, and
most uncertain, pieces of information required to test theoretical predictions
at this time remain 1) the HI column density of the gas, which is only
inferred from the known properties of \ion{Mg}{2}\ absorption systems (either
in QSO or the ones in the outflows of star--forming galaxies); 2) a
characterization of the geometry and spatial extent of the high--column
density gas, both at large scales and in proximity of galaxies; 3) robust
kinematical evidence that the gas is moving in bulk motions and accreting onto
the galaxies.

Luckily, the prospect for progress seems good, even with available technology
and instrumentation. Larger surveys for intervening galaxy
\ion{Mg}{2}\ absorption systems in the background of large cosmic structures
with denser spatial sampling than presented here can be done within current
capabilities by acquiring larger and, to some extent, deeper samples. These
will not only provide a much more robust characterization of the accretion
kinematics, but will also better constrain the geometry and spatial extent of
the gas. Dedicated ultra--deep observations of bright galaxies can even yield
direct estimates of the HI column density from the DLA feature that gas (at
slightly higher redshift than that discussed here) with $N_H>2\times 10^{20}$
cm$^{-2}$ should imprint in the UV spectra of the galaxies. Another distinct
advantage of such an approach is that it offers an independent methodology to
investigate gas inflows with radically different selection bias than the more
traditional one of studying the circum--galactic medium back--illuminated by
the background UV light provided by the galaxy's own stars (Steidel
et~al. 2010). This methodology is effective in characterizing the outflows
from star--forming galaxies. But we do not understand, at present, if and how
the selection bias inherent in such an approach affects the observability of
gas inflows, especially in isolated, field galaxies. Our approach of targeting
the environment of large structures and using redder portions of the UV SED of
star--forming galaxies, where the strong Mg and Fe features are located,
appears promising and worth exploring further.

\acknowledgments

MG, SS, PC and YG acknowledge support from NASA grants HST-GO-9425.36-A,
HST-GO-9822.45-A, and HST-GO-10189.15-A, awarded by the Space Telescope
Science Institute, which is operated by the Association of Universities for
Research in Astronomy, Inc. (AURA) under NASA contract NAS 5-26555. EV
acknowledges support from grant ASI--INAF I/009/10/0. JK acknowledges support
from DFG, via the German--Israeli Project Cooperation grant
STE1869/1-1.GE625/15-1.

\end{document}